\documentclass[aps,pra,twocolumn,showpacs,superscriptaddress,balancelastpage]{revtex4-1}

\usepackage{amsmath}  % AMS Math Package
\usepackage{amsthm}   % Theorem Formatting
\usepackage{amssymb}  % Math symbols such as \mathbb
\usepackage{graphicx} % Allows for eps images
\usepackage{dsfont}
\usepackage{braket}
\usepackage{mathtools}
\usepackage{filecontents}
\usepackage[colorlinks=true, citecolor=blue, linkcolor=blue, urlcolor=blue]{hyperref}

\newcommand{\up}{\uparrow}
\newcommand{\dn}{\downarrow}
\renewcommand{\L}{\mathcal{L}}
\newcommand{\D}{\mathcal{D}}
\newcommand{\E}{\mathcal{E}}
\newcommand{\F}{\mathcal{F}}
\newcommand{\cG}{\mathcal{G}}
\newcommand{\N}{\mathcal{N}}
\newcommand{\C}{\mathds{C}}
\newcommand{\B}{\mathfrak{B}}
\newcommand{\Sc}{\mathcal{S}}
\newcommand{\Ground}{\ket{G}}
\newcommand{\SO}{\mathrm{SO}}
\renewcommand{\O}{\mathcal{O}}
\newcommand{\ketbra}[2][]{|{#2}\rangle_{#1}\langle{#2}|}
\newcommand{\ketbrat}[2]{|{#1}\rangle\langle{#2}|}
\DeclareMathOperator{\trace}{tr}
\newcommand{\vect}[1]{\boldsymbol{#1}}

\DeclareMathOperator{\range}{range}

\DeclareMathOperator{\tr}{tr}
\DeclareMathOperator{\rank}{rank}
\newcommand{\AKLT}{\textrm{AKLT}}
\newcommand{\eff}{\textrm{eff}}
\newcommand{\MP}{\textrm{MP}}
\newcommand{\CW}{\textrm{CW}}
\newcommand{\dc}{\dot{c}}
\newcommand{\I}{\mathds{1}}
\renewcommand{\Re}{\textrm{Re}}
\renewcommand{\Im}{\textrm{Im}}
\newcommand{\GHZ}{\mathrm{GHZ}}
\newcommand{\T}{\vect{T}}

\newtheorem*{lemma*}{Lemma}

\begin{document}
\title{Symmetry-protected dissipative preparation of matrix product states}

\author{Leo Zhou}
\thanks{These authors contributed equally to this work.}
\affiliation{Department of Physics, Harvard University, Cambridge, Massachusetts 02138, USA}

\author{Soonwon Choi}
\thanks{These authors contributed equally to this work.}
\affiliation{Department of Physics, Harvard University, Cambridge, Massachusetts 02138, USA}

\author{Mikhail D. Lukin}
\affiliation{Department of Physics, Harvard University, Cambridge, Massachusetts 02138, USA}

\date{January 20, 2023}

\begin{abstract}
We propose and analyze a method for efficient dissipative preparation of matrix product states that exploits their symmetry properties.
Specifically, we construct an explicit protocol that makes use of driven-dissipative dynamics to prepare a many-body quantum state that features symmetry-protected topological order and non-trivial edge excitations.
The preparation protocol is protected from errors that respect the symmetry, allowing for robust experimental implementation without fine-tuned control.
Numerical simulations show that the preparation time scales polynomially in system size $n$.
Furthermore, we demonstrate that this scaling can be improved to $\mathcal{O}(\log^2 n)$ by using parallel preparation of individual segments and fusing them via quantum feedback.
A concrete scheme using excitation of trapped neutral atoms into Rydberg states via electromagnetically induced transparency is proposed, and generalizations to a broader class of matrix product states are discussed.
\end{abstract}

\maketitle

\section{Introduction}
Entangled many-body states play a central role in understanding strongly correlated quantum matter and constitute the key resource for quantum information science.
Matrix product states (MPSs)\,\cite{MPSreview} form an important class of many-body entangled states that can describe a variety of one-dimensional quantum systems.
Notably, MPSs include states featuring symmetry-protected topological (SPT) order\,\cite{SchuchSPTMPS,Gu2009,ChenGuWen,SenthilSPTReview}, corresponding to exotic quantum phases beyond the standard Landau paradigm of symmetry breaking.
Such states can be prepared either through a sequence of unitary quantum gate operations, or by first engineering the parent Hamiltonian and subsequently preparing its ground state via adiabatic evolution or cooling\,\cite{BollingerTrappedIons, CohenTrappedIonHaldane, BijnenPohlRydbergSpin1, GreinerFermiGas, PlenioDiamond, JunYePolarMolecules, MartinisDigitizedAQC}.
However, generating entanglement among many particles using these approaches is challenging, as it typically requires high-fidelity control of individual interactions while maintaining low entropy for intrinsically out-of-equilibrium systems.
In particular, unavoidable coupling to environment limits the lifetime of these states and hinders their potential applications.

In this paper, we propose and analyze an alternative method to efficiently prepare an MPS by engineering couplings between a system and its environment such that the desired quantum state is obtained as the unique steady state of time evolution.
Such approaches to prepare entangled states have been described previously\,\cite{KrausZoller, Verstraete, TicozziViola2012, ChoBoseKim, AlbertLindbladSymmetries, DissipativeTopological, BlattZollerIonOpenSystem, DissPrepSuperconducting, DissPrepRaoMolmer, DissPrepRK}.
It has also been shown that under certain conditions, the dissipative method outperforms corresponding unitary schemes\,\cite{DissPrepCavities}.
In practice, however, the implementation of these suggested schemes in many-body systems is challenging as it requires engineering of complex interactions and decay channels with environment.
Here, we show how symmetries can be used to design a simple, translation-invariant dissipative process that only requires a single decay channel and global manipulations to create a desired MPS.
Remarkably, similar to symmetry-protected equilibrium phases, this dissipative dynamics is protected from imperfections that respect the symmetry, allowing for robust experimental realizations in large systems with current technologies.
The \emph{symmetry protection} implies that our method does not require high fidelity in controls such as global spin-rotations or interaction strength, which are otherwise essential in conventional unitary schemes.

The paper is organized as follows.
In Sec.~\ref{sec:AKLT}, we describe our method for a well-known example of an MPS that exhibits SPT order. 
We elaborate on how to construct a driven-dissipative dynamics from given symmetry properties of the MPS and rigorously show that the engineered dynamics deterministically prepares the desired states.
Using numerical simulations, we find the state-preparation time scales polynomially with system size $n$.
In Sec.~\ref{sec:parallel}, we show that this scaling of state-preparation time can be further improved to $\O(\log^2 n)$ by first preparing multiple chains in parallel and then connecting them via repeated measurements with feedback.
This corresponds to an exponential improvement over previously known scaling $\O(n^{\log n})$ for generic MPSs~\cite{Verstraete}.
We also provide a detailed analysis of the effect of imperfections in quantum feedback.
We emphasize that our example scheme utilizes the most natural, generic types of environmental couplings as a resource, and hence it can be directly implemented in cold atom systems with existing technologies.
In Sec.~\ref{sec:experiment}, we propose a concrete scheme involving excitation of trapped neutral atoms\,\cite{AtomArray, ParisAtomArray, AtomArrayPhysics, ParisAtomArrayPhysics, LesterAtomArray} into Rydberg states via electromagnetically induced transparency (EIT)\,\cite{REIT}, in which spontaneous emissions of photons from atomic excited states are harnessed as resources.
In Sec.~\ref{sec:generalization}, we generalize our protocol to a broader class of MPSs, including the ground states of all one-dimensional SPT phases, and derive a lower bound on the number of required decay channels that may be saturated by an explicit construction.
We discuss our results and outlook in Sec.~\ref{sec:outlook}.

\section{Dissipative Preparation of AKLT states\label{sec:AKLT}}

\subsection{Ground states of AKLT Hamiltonian}
We illustrate our scheme by starting with an example to deterministically prepare a chain of spin-1 particles into the ground states of a gapped, frustration-free Hamiltonian
\begin{equation}
H_\textrm{AKLT} = \sum_i[ \vec{S}_i \cdot \vec{S}_{i+1} + \frac{1}{3} (\vec{S}_i \cdot \vec{S}_{i+1} )^2],
\end{equation}
where $\vec{S}_i$ is the spin-1 vector operator acting on a particle at site $i$.
First studied by Affleck, Kennedy, Lieb, and Tasaki (AKLT), the ground states of $H_\textrm{AKLT}$ are paradigmatic examples of MPSs and model states for the Haldane phase\,\cite{AKLT,Haldane1983,HaldaneSPT}.
While under periodic boundary condition, $H_\textrm{AKLT}$ has a unique ground state; under open boundary condition, the ground states are fourfold degenerate due to two fractionalized degrees of freedom on the edges.
These constitute a signature of symmetry-protected topological order, which can be experimentally verified by measuring a non-local string order parameter\,\cite{StringOrder,StringOrder2}.

The ground states of AKLT Hamiltonian have exact MPS representations~\cite{MPSreview}.
For a system of $n$ spin-1 particles, the unnormalized AKLT ground states can be written as
\begin{equation}
\ket{G_{ab}^n} = \sum_{\{s_i\}} \braket{a|A^{(s_1)}A^{(s_2)}\cdots A^{(s_n)}|b}\ket{s_1s_2\cdots s_n},
\end{equation}
where $s_i\in\{\pm1, 0 \}$ runs over three possible spin projections along the $\hat{z}$-axis for a particle at site $i$.
The quantum amplitude for each many-body basis state $\ket{s_1s_2 \dots s_n}$ is obtained from the products of $D\times D$ matrices $A^{(s)}$, and the boundary conditions for the matrix products are specified by a row (column) vector $\bra{a}$ ($\ket{b}$) of dimension $D$.
Specifically, for AKLT ground states, $A^{(s)}$ can be concisely written using ($D$\,=\,2) Pauli matrices $\sigma^x$, $\sigma^y$, $\sigma^z$, and $\sigma^{\pm} = (\sigma^x \pm i \sigma^y)/2$:
\begin{align}
A^{(1)} = \sqrt{\frac23}\sigma^+, ~
A^{(0)} = -\sqrt{\frac13}\sigma^z, ~
A^{(-1)} = -\sqrt{\frac23}\sigma^- .
\end{align}
The choices of vectors $a,b \in \{\uparrow, \downarrow\}$ distinguish the four  degenerate ground states with different fractionalized edge states under open boundary conditions~\footnote{We note that our convention here for labeling the edge states $a,b$ is different from the convention where the AKLT state is understood as projecting pairs of virtual spin-1/2 particles in singlet bonds back into the spin-1 particles, up to a basis change by $\mathds{1}\otimes i\sigma^y$.}.
Under periodic boundary condition, the unique ground state is $\ket{G^n_\circ} \equiv \sum_a \ket{G_{aa}^n}$.

We note a few properties of these AKLT states.
With the notation $\ket{G^n_{ab}}$,
one can conveniently rewrite the quantum state of an $n$-particle system as a linear superposition of composite systems, each with $m$ and $n-m$ particles, i.e.
$\ket{G_{ab}^n}=\sum_c \ket{G_{ac}^m}\ket{G_{cb}^{n-m}}$.
Moreover, the overlap between two AKLT states with different edge states can be analytically evaluated;
using the transfer matrix $\T=\sum_{s} A^{(s)*}\otimes A^{(s)}$, we obtain exponentially small overlaps between distinct states with a normalization factor $1/2$:
\begin{align}
\label{eq:AKLToverlap}
&\braket{G_{ab}^n|G_{a'b'}^n} = \braket{aa'|\T^n|bb'} =  \nonumber\\
&\qquad \frac{1}{2} \delta_{aa'}\delta_{bb'} (1 - (-1)^{\delta_{ab}} \epsilon^{n}) +\delta_{ab}\delta_{a'b'} (1-\delta_{aa'})\epsilon^n,
\end{align}
where $\epsilon = -1/3$. 
Finally, the AKLT ground state exhibits topological order that is protected by the symmetry groups $D_2$ (dihedral group corresponding to permutation of spin axes), $\mathcal{T}$ (time-reversal symmetry), and $\mathcal{P}$ (bond-inversion symmetry). 
While the presence of any of these symmetries can protect the non-trivial order reflected in the double degeneracy in the entanglement spectrum, the $D_2$ symmetry is necessary and sufficient to protect the string order parameter~\cite{HaldaneSPT}.
In addition to $D_2$, the parent Hamiltonian $H_{\AKLT}$ respects a larger symmetry group of $\SO(3)$, corresponding to global rotation of spins.

\subsection{Constructing driven-dissipative dynamics}

Our key idea is to use $\SO(3)$ symmetry of the parent Hamiltonian $H_\textrm{AKLT}$ for preparation of an exact AKLT ground state.
By converting energy penalties imposed by $H_\textrm{AKLT}$ into dissipative penalties in the form of decay channels, we can engineer a process that effectively cools to the ground states.
More specifically, we start with a dissipative dynamics that eliminates one type of excitation in $H_\textrm{AKLT}$.
Then, all other types of excitations can be eliminated using global spin-rotations in $\SO(3)$.
Since spin-rotations are symmetries of $H_\textrm{AKLT}$, their implementations are robust against imperfections in control parameters such as durations, phases, or strengths of electromagnetic driving.

We consider a Markovian driven-dissipative dynamics described by a quantum master equation:
\begin{align}
\dot\rho = \L\rho \equiv -i[H,\rho] + \sum_\mu \Gamma_\mu \D[c_\mu]  \rho,
\end{align}
where $\rho$ is the density operator of a system, $H$ is a Hamiltonian governing coherent dynamics, and $\D[c_\mu]\rho \equiv c_\mu \rho c_\mu^\dagger - \{c_\mu^\dagger c_\mu, \rho\} /2$ characterizes incoherent dynamics by jump operators (i.e., decay channel) $c_\mu$ at rate $\Gamma_\mu$.
We can interpret the dynamics of $\L$ as the system evolving with a non-Hermitian Hamiltonian $H_\text{eff}$\,=\,$H-i\sum_\mu\Gamma_\mu c_\mu^\dag c_\mu/2$, while stochastically undergoing quantum jumps $\rho$\,$\mapsto$\,$c_\mu \rho c_\mu^\dagger$ at rates $\trace(\Gamma_\mu c_\mu^\dagger c_\mu \rho)$ for each jump operator\,\cite{gardiner2004quantum}.

In order to construct a simplest possible $\L$ that prepares an AKLT ground state,
we exploit the $\SO(3)$ symmetry that conserves total angular momentum.
In particular, each term in $H_\textrm{AKLT}$ can be written as \mbox{$2 P_i - 2/3$}, where $P_i$ is the projection operator onto the subspace of total angular momentum $J_i$\,=\,$S_i + S_{i+1}$\,$=$\,2 for the pair of particles $(i,i+1)$.
Hence, a state $\ket{G}$ minimizes the energy if it has no population in the $J_i$\,=\,2 manifold, i.e. $P_i\ket{G}$\,=\,$0$ for every nearest-neighboring pair.
Under open boundary condition, there are four such states $\ket{G_{ab}}$, labelled by two spin-1/2 edge degrees of freedom $a,b$\,$\in$\,$\{ \uparrow, \downarrow\}$.
Under periodic boundary condition, only a unique state $\ket{G_\circ}$\,$\propto$\,$\ket{G_{\uparrow \uparrow}} + \ket{G_{\downarrow \downarrow}}$ satisfies the constraints.
Below, we use $\Ground$ to denote the ground state(s) when boundary conditions are not specified.

To prepare $\Ground$, we use jump operators to depopulate $J_i$\,=\,2 manifold of every neighboring pair.
For example, we can set $H$\,=\,0 and use five types of jump operators,
$c_{m}^{(i)}$\,=\,$\ketbrat{\phi_m}{J=2,J^z=m}_{i,i+1}$,
where $\{|J$\,=\,$2,J^z$\,=\,$m\rangle_{i,i+1}$\,:\,\,$m$\,=\,$-2,\dots, 2\}$ is an orthonormal basis spanning the $J_i$\,$=$\,$2$ manifold for the pair of spins $(i,i+1)$, and $\ket{\phi_m}$ is any other quantum state with nonzero population in $J_i$\,=\,$0,1$ manifolds ($\bra{\phi_m} P_i \ket{\phi_m}$\,$<$\,1).
With these jump operators, quantum jumps occur at rate $\Gamma_\textrm{total}$\,=\,$\Gamma \sum_{i,m} \trace{(\rho c_{m}^{(i)\dag} c_{m}^{(i)})}$\,=\,$\Gamma \sum_i \trace{(\rho P_i)}$, which vanishes only for the ground state $\Ground$.
This implies that $\Ground$ is a steady state of $\L$, and any other quantum state will undergo a series of quantum jumps.

Using $\SO(3)$ symmetry, this construction can be effectively realized with only one type of jump operator via global coherent manipulations $H$.
More specifically, let us consider a dynamics with only one jump operator, $c_{2}$\,=\,$\ket{00}\langle J$=$2,J^z$=$2|$\,=\,$\ketbrat{00}{++}$, written in the $S^z$ basis $\{\ket{+},\ket{0},\ket{-}\}$.
After time evolution over duration $\tau/5$, we apply a fast global pulse $V$\,$=$\,$\exp[i (2\pi/5) \sum_i S_i^y]$, rotating the entire spin ensemble by an angle $2\pi/5$ about the $y$-axis.
In a rotating frame, this operation implements the jump operator $V^\dag c_{2} V$.
Repeated multiple times, we obtain five distinct jump operators $\bar{c}_{\nu}$\,$\equiv$\,$(V^\dag)^\nu c_{2} (V)^\nu$ for $\nu$\,$\in$\,$\{0,\dots, 4\}$ after the $\nu$-th (modulo 5) pulses.
For a sufficiently short $\tau$\,$\ll$\,$1/\Gamma$, the effective Liouvillian of the five-pulse cycle can be well approximated using leading-order Magnus expansion by
\begin{align}
\L_\textrm{MP} = (\Gamma/5) \sum_{i}\sum_{\nu=0}^4\D[\bar{c}_\nu^{(i)}].
\end{align}
Note that the purpose of global rotations is to use a single jump operator for depopulating different states; different choices of angles and axes are equally effective as long as states rotated from $\ket{++}$ span the entire $J$\,=\,2 manifold.
We may also employ a time-independent Hamiltonian $H_\textrm{CW}$\,=\,$\omega \sum_i S_i^y$ to continuously rotate the ensemble, leading to an effective Liouvillian
\begin{align}
\L_\text{CW} = \frac{\omega}{2\pi}\int_0^{2\pi/\omega}dt\,\Gamma
\sum_i \D[ e^{iH_\textrm{CW}t}c_{2}^{(i)} e^{-iH_\textrm{CW} t} ].
\end{align}
In both cases, the corresponding quantum jump rates vanish if and only if the system is in $\Ground$.

\begin{figure}
  \centering
  \includegraphics[width=0.5\textwidth]{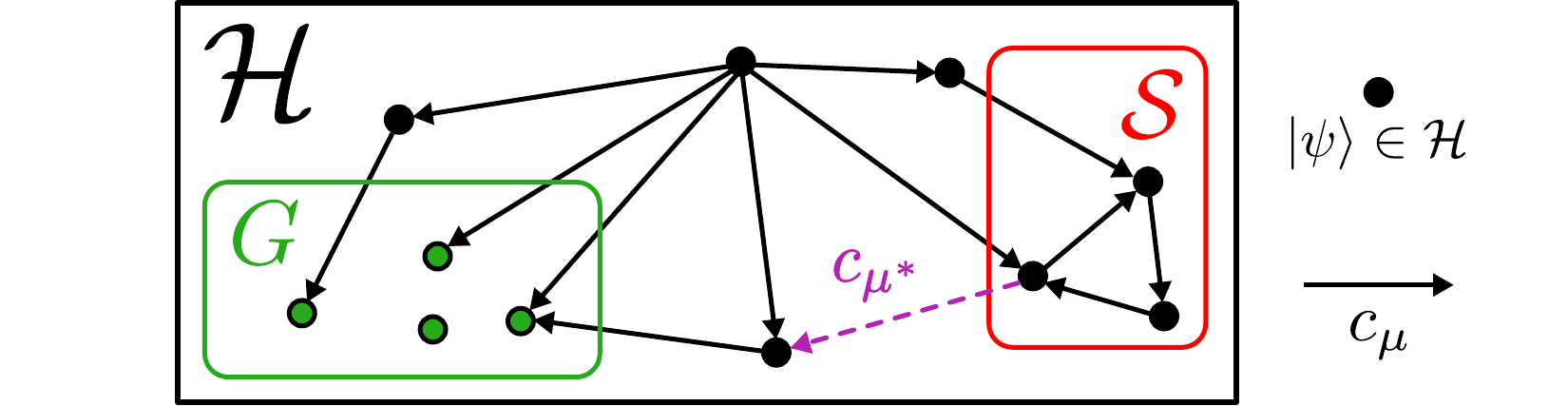}
  \caption{\label{fig:steadystates}Visualization of incoherent quantum jumps as random walks on a directed graph in Hilbert space $\mathcal{H}$. Here $G$ is the subspace of steady states that do not undergo quantum jumps.
  In the absence of the dashed arrow $c_{\mu^*}$, the subspace of three states $\Sc$ is closed under quantum jumps, allowing a mixed steady state to form.
  The presence of $c_{\mu^*}$ eliminates this possibility.}
\end{figure}

\subsection{Proving uniqueness of steady states\label{sec:uniqunessAKLT}}
While our construction of $\L_\textrm{MP}$ and $\L_\textrm{CW}$ ensures that $\Ground$ is a steady state, one can imagine an undesired mixed steady state that forms in dynamical equilibrium from the combination of coherent evolution and incoherent quantum jumps (Fig.\,\ref{fig:steadystates}).
Such mixed steady states may arise only if there exists a subspace $\Sc$ orthogonal to $\Ground$ and closed under jump operators, $c_\mu \Sc$\,$\subseteq$\,$\Sc$\,\cite{KrausZoller}.
Physically, this means that states in $\Sc$ cannot reach $\Ground$ even with arbitrarily many applications of jump operators $c_\mu$, allowing an equilibrium to form by their mixtures.
In our scheme, we prove the following lemma that guarantees that the desired state $\ket{G}$ is the unique steady state:

\begin{lemma*}
For any finite system with size $n$\,$\ge$\,$2$ under open boundary condition, all states can reach $\Ground$ with some application of jump operators in $\L_\textnormal{MP}$ or $\L_\textnormal{CW}$, implying $\Ground$ is the unique steady state.
\end{lemma*}

\begin{figure*}
  \centering
  \includegraphics[width=\textwidth]{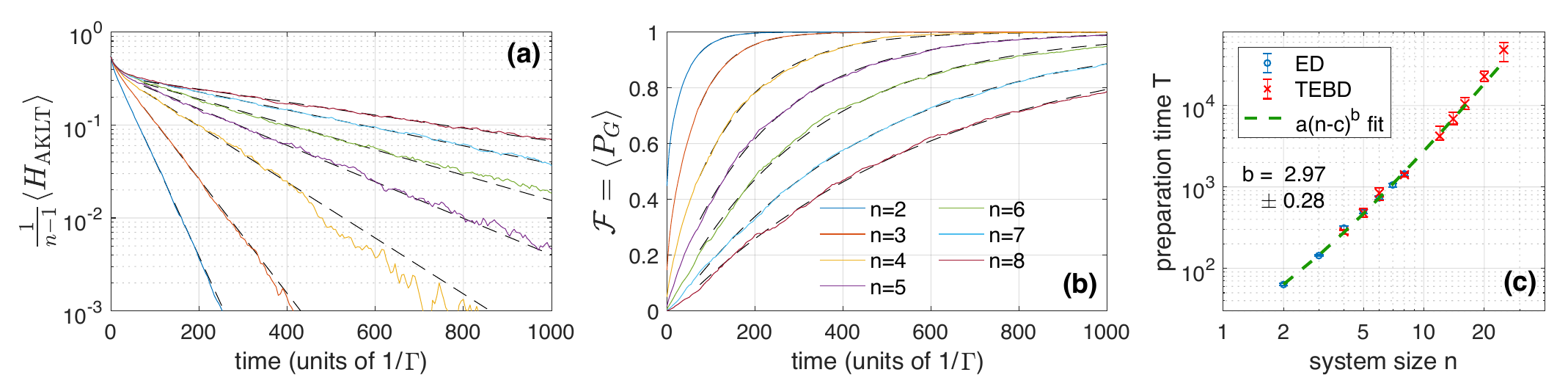}
  \caption{(a) and (b) Numerical simulation of $\L_\text{MP}$ for $\Gamma$\,$=$\,$1$ with a maximally mixed initial state, using exact diagonalization (ED) for system size up to $n$\,$=$\,$8$. Both the energy density $\braket{H_{\rm AKLT}}/(n-1)$ and the state-preparation fidelity $\F$ in the long-time regime are fitted to an exponential function (dashed lines). (c) A log-log plot of the fitted preparation time to achieve $\F$\,$=$\,$0.9$ from simulations using ED and TEBD algorithms, as a function of system size, up to $n$\,$=$\,$25$. Error bars are 90\% confidence intervals.}\label{fig:numerics}
\end{figure*}

We now sketch the proof of this Lemma for open boundary condition, where the four $\ket{G_{ab}}$ states are unique steady states;
more details on the proof can be found in Appendix~\ref{sec:proof_uniqueness}.
The proof uses induction on system size $n$.
Note that under an open boundary condition, the four $\ket{G_{ab}}$ states are the desired steady states.
Let us define $D_n = \text{span}\{\ket{G_{ab}^{n}}\}$ as the subspace of AKLT ground states of $n$ spins.
For $n$\,=\,$2$ and 3, the uniqueness of steady states can be checked by exact diagonalization.
Our induction hypothesis is that any $n$-spin input state can reach $D_n$ with some applications of jump operators, via some polynomial function $f_{[n]}(\{\bar{c}_\nu\})$ (which may depend on the input state).
For the sake of contradiction, let us assume that a state $\ket{\psi^{n+1}}$ cannot reach $D_{n+1}$ with any sequence of jump operators $\bar{c}_\nu^{(i)}$ in $\L_\textrm{MP}$.
(The same argument holds for $\L_\textrm{CW}$.)
We will then construct a sequence of jump operators, involving some $f_{[n]}$ on the first $n$ spins followed by some $\bar{c}_\nu^{(n)}$ acting on spin $n$ and $n+1$, so that  $\bar{c}_\nu^{(n)} f_{[n]} \ket{\psi^{n+1}}$ reaches $D_{n+1}$, leading to a contradiction.

To begin, we know that by our induction hypothesis, there exists $f_{[n]}(\{\bar{c}_\nu\})$ so that $f_{[n]} \ket{\psi^{n+1}}$ has nonzero population in $D_n$.
Since the AKLT Hamiltonian is frustration-free, $f_{[n]} \ket{\psi^{n+1}}$ must also have nonzero population in $D_{n-1}$, i.e. the AKLT ground states on the first $n-1$ spins.
We then do a general decomposition of $f_{[n]}\ket{\psi^{n+1}} = \ket{\phi} + \ket{\phi^\perp_1} + \ket{\phi^\perp_2}$, where
\begin{align}
\ket{\phi} = \sum_{a,b=\dn}^\up \sum_{z=-2}^2 \phi_{abz}\ket{G_{ab}^{n-1}} \ket{z}.
\end{align}
Here $\ket{z}$ runs through the 5 states in the $J=2$ manifold on the last two spins. The remaining parts of the wavefunction $f_{[n]}\ket{\psi^{n+1}}$ are
\begin{align}
\ket{\phi^\perp_1} &= \sum_{abst} \phi_{abst}^\perp \ket{G_{ab}^{n-1}}\ket{G_{st}^2}, \\
\ket{\phi^\perp_2} &=\sum_{\mu s} \phi_{\mu s}^\perp \ket{E_\mu^{n-1}}\ket{s},
\end{align}
where $\ket{E_\mu^{n-1}}$ runs through all the excited eigenstates of the $H_\AKLT$ on the first $n-1$ spins, and $s$ runs through all 9 possible 2-spin states.
We now consider two cases:

{\it Case (i)}: Suppose $\phi_{abz} = 0$. Then $f_{[n]}\ket{\psi^{n+1}} = \ket{\phi^\perp_1} + \ket{\phi^\perp_2}$.
From our inductive hypothesis, $f_{[n]}\ket{\psi^{n+1}}$ must have nonzero population in $D_n$ and $D_{n-1}$; this can only come from the $\ket{\phi^\perp_1}$ part, and thus there must be nonzero coefficients $\phi_{abst}^\perp\neq 0$.
On the other hand, since we have assumed that $\ket{\psi^{n+1}}$ cannot reach $D_{n+1}$, we must have
$0= \braket{G_{pq}^{n+1}|\phi^\perp_1}$, which is only possible if  $\ket{\phi^\perp_1}$ has the state of spins at $n-1$ and $n$-th sites in the $J$\,=\,$2$ manifold.
However, that would then imply that $\ket{\phi^\perp_1}$ has zero population in $D_n$, contradicting our inductive hypothesis.

{\it Case (ii)}: Now suppose $\phi_{abz}\neq 0$ for some $a,b,z$. 
Note that for any jump operator $\bar{c}_\nu^{(n)}$ acting on spins $n$ and $n+1$, 
we have $\braket{G_{pq}^{n+1}| \bar{c}_\nu^{(n)} | \phi^\perp_j} = 0$ for $j=1, 2$, since $\bar{c}_\nu\ket{G_{st}^2} = 0$ and $\braket{G_{ab}^{n-1}|E_\mu^{n-1}}=0$.
Then our assumption that $\ket{\psi^{n+1}}$ cannot reach $D_{n+1}$ gives the following set of linear equations for $\phi_{abz}$:
\begin{align}
0 &= \braket{G_{pq}^{n+1}| \bar{c}_\nu^{(n)} f_{[n]} | \psi^{n+1}} \nonumber \\
&= \sum_{abz} \phi_{abz}\braket{G_{pq}^{n+1}| \bar{c}_\nu^{(n)} |G_{ab}^{n-1}}\ket{z}
\quad \forall p,q,\nu.
\label{eq:uniqueness}
\end{align}
Since $\ket{G^n_{ab}}$ and $\ket{G^{n+1}_{pq}}$ have explicit MPS formulas, one can analytically compute these expressions and find that only the trivial solution $\phi_{abz}$\,=\,$0$ are allowed for $n\ge 3$.
This yields a contradiction and implies that all states $\ket{\psi^{n+1}}$ can reach at least one of the states $\ket{G^{n+1}_{pq}}$ in $D_{n+1}$ with some application of jump operators.

\subsection{Numerical simulations and scaling}
We numerically study the efficiency of our protocol via stochastic wavefunction method for systems of up to $n=25$ particles.
We use both exact diagonalization (for $n$\,$\le$\,$8$) and time-evolving block decimation (TEBD) algorithm\,\cite{Vidal,*Vidal2} in MPS representations (for $n$\,$\le$\,$25$);
more details are discussed in Appendix~\ref{sec:detail_numerics}.
We initialize the system in a random product state (representing a maximally mixed state), and evolve under $\L_\mathrm{MP}$ with open boundary condition.
We then monitor the energy density with respect to $H_\text{AKLT}$, as well as the fidelity of state preparation $\F$\,$=$\,$\braket{P_{G}}$, where $P_G$ is the projector onto the ground states.
The results in Figs.\,\ref{fig:numerics}(a) and (b) demonstrate that both observables exponentially converge to their corresponding values for AKLT states in all system sizes.
We extract the state-preparation time $T$ by first fitting $1-\F$ to an exponential in the long-time regime and extrapolating $\F (T)$\,$=$\,$0.9$.
We find that $T$ generally increases with system size $n$.
Plotted as a function of $n$ [Fig.\,\ref{fig:numerics}(c)], we find a polynomial scaling $T$\,$\sim$\,$\O(n^{2.97})$.
This scaling is consistent with the more complicated protocol in Ref.\,\cite{Verstraete} that requires up to $O(n^{\log_2 n})$ time, up to system sizes simulated in this work.

\section{Improving Scaling via Parallelization and Quantum Feedback\label{sec:parallel}}

\begin{figure}
\includegraphics[width=\linewidth]{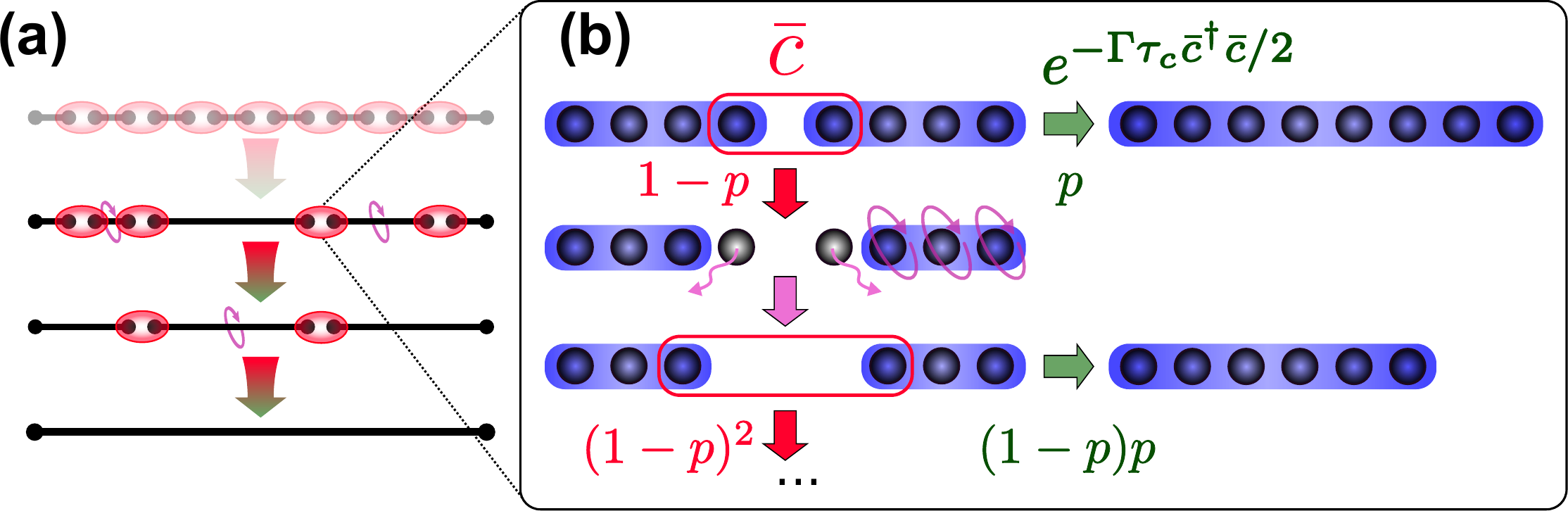}
\caption{(a) Scheme for preparing AKLT states in parallel to achieve logarithmic scaling. Many short chains of spins in AKLT states are prepared initially, and adjacent chains are connected probabilistically in parallel.
Failures are addressed with quantum feedback on every other segments before re-attempting the connections.
(b) Illustration of the connection algorithm, where the success of each attempt occurs with probability $p$, while failure can be corrected by discarding two spins and reattempting. Note that only one success is necessary among all the attempts. On average, $1/p$ attempts are sufficient to obtain a successful connection, with a constant overhead of $2(1-p)/p$ spins.}
\label{fig:connect}
\end{figure}

While the time our protocol needs to prepare AKLT states is already shown to have an efficient polynomial scaling from our numerical simulations, we now provide a method to exponentially improve this scaling to $\O(\log^2 n)$.
Similar to approaches used in quantum repeaters\,\cite{QuantumRepeater}, this exponential speedup is possible by preparing multiple chains in parallel, which are subsequently connected into a single long chain [Fig.\,\ref{fig:connect}(a)].
The key ingredient is the ability to efficiently connect or fuse two AKLT chains into a single entangled state [Fig.\,\ref{fig:connect}(b)], which we now describe.

\subsection{Connecting two AKLT chains\label{sec:connectAKLT}}
Suppose we have independently prepared two chains of $m$ spins in AKLT states.
As they are initially unentangled, their state can be written as $\ket{\psi_0} = \N_0 (\sum_{b}\alpha_b \ket{G^m_{ab}}) \otimes (\sum_c \beta_c \ket{G^m_{cd}})$, where $\vec\alpha,\vec\beta\in\C^2$ characterizes the edge states at the interface, and $\N_0$ is some normalization constant.
The edge states represented by the indices $a$ and $b$ are unimportant for the connection.
To connect the two chains into one, we can turn on the jump operators $\{\bar{c}_{\alpha}^{(m)}\}$ acting on spins $m$ and $m+1$ at the interface, and evolve for some time $\tau_c$.
Then, we monitor quantum jump events to determine if we have succeeded in creating an AKLT state $\ket{\psi_f^{2m}}$\,$=$\,$\N_f \ket{G_{ad}^{2m}}$ with doubled length ($\N_f$ is some normalization constant).

A successful connection is heralded by the absence of quantum jumps, in which case evolution under the non-Hermitian Hamiltonian guides the system into an AKLT state of the combined chain.
For sufficiently long $\tau_c$, the success probability is given by the overlap between initial and desired states (see Appx.~\ref{sec:quantum_jump}).
This can be computed using Eq.~\eqref{eq:AKLToverlap} to be
\begin{equation} \label{eq:succ-prob}
p = \left|\braket{\psi_f^{2m}|\psi_0}\right|^2  = \frac{|\vec\alpha\cdot\vec\beta|^2}{2|\vec\alpha|^2|\vec\beta|^2} + \O(3^{-m}).
\end{equation}
Note that if the edge states $\vec{\alpha}$ and $\vec\beta$ are random vectors in $\C^2$, we have on average $p\simeq 1/4$.
When the edge states are aligned, i.e. $\vec\alpha \propto \vec\beta^*$, we obtain the maximum success probability of $p_\text{max}=1/2$.
The resultant state has an exponentially small error $\epsilon$\,$\le$\,$e^{-\O(\tau_c)}$.

The failure of the connection is signaled by detection of a quantum jump $\bar{c}_\alpha$, after which the state of the system changes according to $\ket{\psi_0} \mapsto \bar{c}_{\alpha}^{(m)} \ket{\psi_0}$.
In this case, one can discard the pair of spins $(m,m+1)$ and then attempt the connection procedure again with two chains of length $m-1$.
However, it turns out that quantum jumps affect the success probability of subsequent connection attempts, which in fact vanishes for this protocol without additional intervention, a phenomenon that we will explain in the following paragraph.
Nevertheless, we can restore the success probability to $p_{\max}=1/2$ by applying a global spin rotations $U$\,=\,$(e^{i \pi S^y})^{\otimes{m-1}}$ to one of the chains.
This quantum feedback makes the procedure very efficient, since multiple repeated failures are exponentially unlikely and only one success is sufficient to fuse two chains.
The number of attempts necessary follows the geometric distribution, and on average we need $(1-p)/p$ attempts with the loss of $2(1-p)/p$ particles per connection.
By performing these connection procedures in parallel, we can quickly prepare an AKLT state of $n$ spins in $\O(\log^2 n)$ time, as we show in Sec.~\ref{sec:parallelscaling}.

We now explain why a single failed connection attempt will cause subsequent attempts to fail, unless appropriate quantum feedback is applied.
It turns out that this problem occurs whenever the matrix product state we want to prepare respects bond-inversion symmetry, but here we first focus on the example of AKLT states which has an intuitive explanation (see Sec.~\ref{sec:general-parallel} for the general case).
Note that we can interpret the dynamics under $\L=\D[\bar{c}_\theta]$ as a continuous measurement of whether the pair of spins has total angular momentum $J_{\theta}=+2$, where $J_{\theta}=e^{-i\theta J_y} J_z e^{i\theta J_y}$.
To be more specific, consider four spin-1 particles $\vec{S}_1,\vec{S}_2,\vec{S}_3, \vec{S}_4$, and imagine that we are performing a connection between spins 2 and 3 by continuously measuring $\vec{J}=\vec{S}_2+\vec{S}_3$.
Suppose we decompose each spin-1 into two virtual spin-$\frac12$ particles: $\vec{S}_i = \vec{s}_{i,L}+\vec{s}_{i,R}$.
It is known~\cite{AKLT} that an AKLT state can be constructed by starting with singlet bonds of virtual spin-$\frac12$ particles where $s_{i,R}+s_{i+1,L}=0$ for all $i$, and then projecting back into the triplet subspace of the original pairs of virtual spin-$\frac12$ particles where $s_{i,L}+s_{i,R}=1$.
The detection of a quantum jump $c_\theta$ in a failed connection attempt implies that $J_\theta = 2$, which is only possible if $s_{2L}^\theta = s_{2R}^\theta=s_{3L}^\theta=s_{3R}^\theta=+\frac12$.
Due to the singlet bond conditions, this automatically implies that $s_{1R}^\theta = s_{4L}^\theta=-\frac12$.
Subsequently, when we retry the connection with spins 1 and 4 (after discarding 2 and 3), the two virtual spin-$\frac12$ particles at the interface are in the state $\ket{s_{1R}^\theta=-\frac12}\ket{s_{4L}^\theta=-\frac12}$, which has no overlap with the desired singlet bond state $\ket{+\frac12}\ket{-\frac12}-\ket{-\frac12}\ket{+\frac12}$.
Hence the overlap with the AKLT state is zero, and the connection is certain to fail.
Now, applying $U$\,=\,$(e^{i \pi S^y})^{\otimes{m-1}}$ to the first chain flips $s_{1R}^\theta$ so that the resultant state is $\ket{+\frac12}\ket{-\frac12}$.
This has an overlap of 1/2 with the singlet state, restoring our success probability to roughly 1/2.

\subsection{Scaling of preparation time of the parallelized protocol\label{sec:parallelscaling}}

We are now ready to describe and analyze the full parallelized protocol to prepare AKLT states on large system sizes $n$ in $\O(\log^2 n)$ time.
To prepare a length-$n$ chain, we first prepare $\O(n/n_0)$ AKLT chains of length $n_0$, and then apply the above connection procedure in parallel for adjacent chains, as illustrated in Fig.~\ref{fig:connect}(a).
Since on average we lose $n_c=2(1-p)/p$ particles per connection, we should choose $n_0 \gg n_c$, and prepare $N_{\rm chains} = (n-n_c)/(n_0-n_c)$ such chains.
We can imagine attempting the connection simultaneously for all junctions between chains, and we will typically have some successes and some failures.
By noting the locations of the failures and applying the quantum feedback $e^{i\pi S^y}$ to every other segment, we can then re-attempt connections on the junctions that have failed, which will then succeed with $p\simeq 1/2$ as shown above.
This process is repeated for multiple rounds until all junctions become connected, and the system becomes one connected chain.
The successful connection of each junction occurs independently and probabilistically; hence their order may be arbitrary.
The probability of completing all  $N_{\rm chains} -1 $ connections after $K$ rounds is
\begin{align}
\Pr\{\textrm{completion}\} =  (1-(1-p)^K)^{N_{\rm chain}-1}.
\end{align}
To achieve a completion probability of $p_{\rm comp}$, we need $K_*=\log[1-p_{\rm comp}^{1/(N_{\rm chain}-1)}]/\log(1-p) $\,=\,$ \O(\log n)$ rounds. 
Hence, the time required to successfully complete all connection with constant probability and obtain a length-$n$ AKLT chain is 
\begin{equation}
    T(n)  \le T_0 + K_* \tau_c  +  (K_*-1)\tau_r,
\end{equation}
where $T_0$ is the preparation time of the length-$n_0$ chains, $\tau_c$ is the time for each connection attempt, and $\tau_r$ is the time required for feedback after each failed attempt.
Recall that each successful connection induces an error of $\epsilon$\,$\le$\,$e^{-\O(\tau_c)}$ in the quantum state.
Thus, a total of $\O(n/n_0)$ connections yield a final error of $\E\le \O(n/n_0)e^{-\O(\tau_c)}$, which means we should choose $\tau_c$\,$=$\,$\O(\ln(n/n_0\E))$ to achieve a final error of $\E$.
Assuming arbitrarily fast classical communication and control, the quantum feedback of applying homogeneous spin-rotation $e^{i\pi S_y}$ to a subset of the chains can be done in a system-size independent time $\tau_r=\O(1)$.
Thus, the average time necessary to prepare an AKLT state of length $n$ with bounded error $\E$ in this parallelized protocol is
\begin{equation}
T(n) = \O(\log(n)\log (n/\E)) = \O(\log^2 n).
\end{equation}
The required number of spins that we need initially is  $n_0(n-n_c)/(n_0-n_c)$, indicating that an $\O(n)$ (i.e., constant factor) spatial overhead is sufficient.

While so far we have assumed that we can detect the occurrence of quantum jumps perfectly, this parallelized protocol remains very efficient even when the detection is imperfect.
We describe two possible methods to address detection inefficiency.
The first method is to use jump operators of the form $c=\ketbra{++}$ for connection, which will cause an indefinite number of quantum jumps to occur once the first quantum jump occurs, enhancing the quantum jump signal.
Alternatively, we can also slowly turn on additional jump operators near the interface if we do not detect any quantum jump initially; this serves to confirm that the two original chains have been successfully connected, as any failed connection attempt that evaded detection would cause more quantum jumps.
In both cases, the scaling of our parallelized protocol is not significantly altered compared to the ideal case, and can be largely accounted for by modifying the effective success probability $p$.
We provide a more detailed analysis of these two methods in Appendix~\ref{sec:imperfectQJdetect}.

\section{Experimental realization\label{sec:experiment}}

The key task in implementing our protocol is to engineer the nearest-neighbor jump operators.
Such engineering has been previously demonstrated in systems of trapped ions\,\cite{BlattZollerIonOpenSystem}.
Here, we provide an explicit method to realize our scheme in systems of trapped atoms\,\cite{AtomArray,ParisAtomArray,AtomArrayPhysics,ParisAtomArrayPhysics,LesterAtomArray} based on the Rydberg-EIT mechanism\,\cite{REIT}.
We consider a five-level system consisting of a metastable Rydberg state $\ket{r}$, a short-lived excited state $\ket{e}$, and three long-lived hyperfine ground states $\ket{+},\ket{0}$, and $\ket{-}$ as shown in Fig.\,\ref{fig:}(a).
Using lasers, we coherently couple the ground state $\ket{+}$ to the excited state with a time-dependent Rabi frequency $g(t)$.
The excited state is further coupled to the Rydberg state with Rabi frequency $\Omega$.
Owing to large dipole moments, simultaneous excitations of two Rydberg states within distance $R$ are suppressed by an interaction energy shift that decays as $1/R^6$\,\cite{DipoleBlockade}.

\begin{figure}[tb]
\centering
\includegraphics[width=0.48\textwidth]{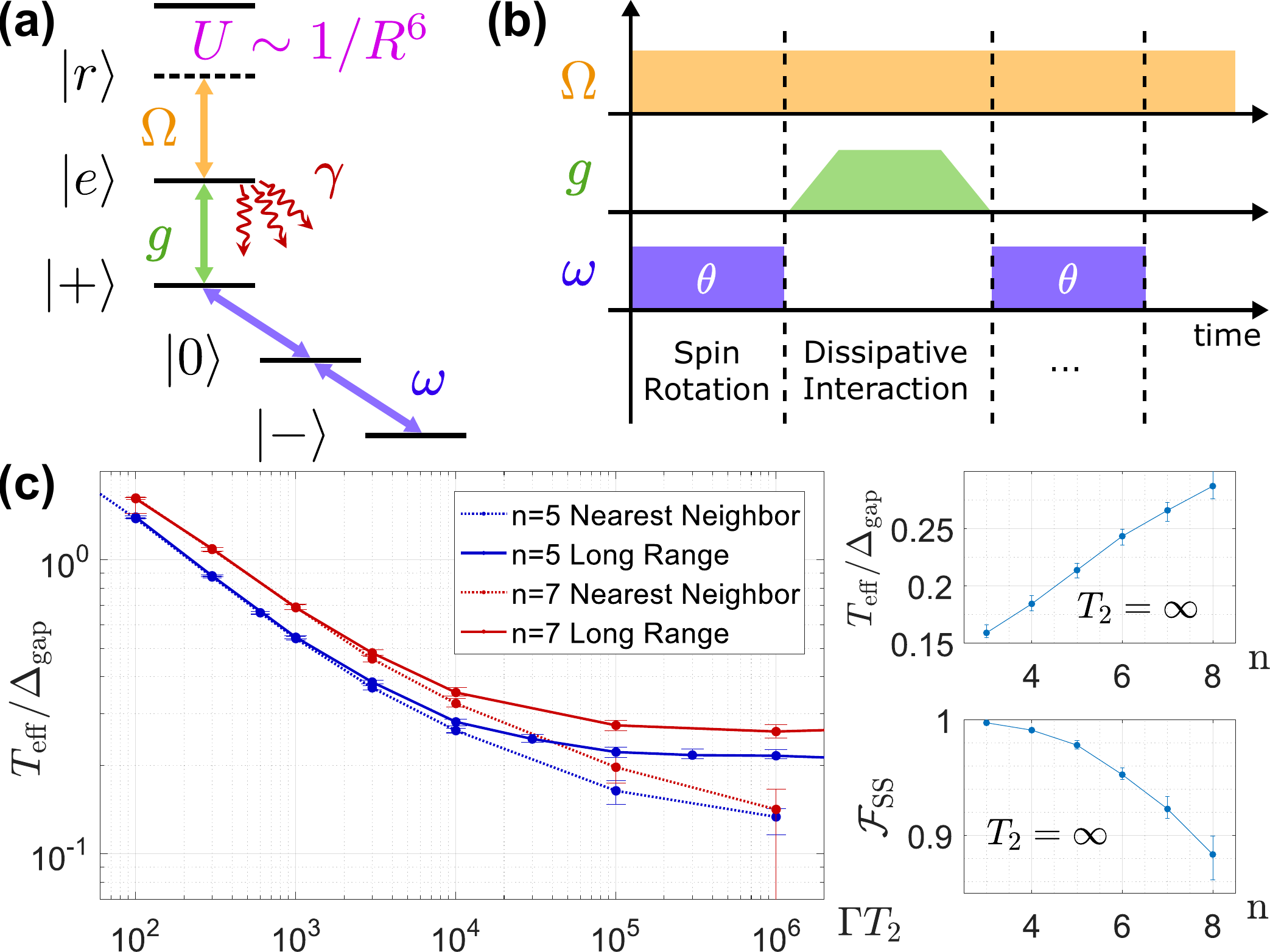}
\caption{\label{fig:EITimpl}(a) Atomic level diagrams for Rydberg-EIT implementation of our jump operators, where $\ket{r}$ is a Rydberg level with strong interaction, and $\ket{e}$ is an excited level with short lifetime $1/\gamma$. The lower three levels encode the spin-1 particles.
(b) Pulse sequence to engineer $\L_\text{MP}$.
(c) Effects of the finite dephasing time ($T_2$) of spin-1 levels and long-range interaction. We use steady state fidelity $\F_\textrm{SS}$ from numerical simulations to calculate effective temperature $T_\textrm{eff}$ in units of $\Delta_\textrm{gap}$, the energy gap of $H_\textrm{AKLT}$.}
\end{figure}

In the absence of interactions, our coherent driving ensures that every atom supports three stable states $\ket{-}, \ket{0}$, and $\ket{D(t)}$\,$\propto$\,$\Omega \ket{+} - g(t)\ket{r}$ for arbitrary choices of $g$ and $\Omega$.
We use these three states to encode the spin-1 degree of freedom.
When $g(t)$ slowly increases starting from zero, $\ket{+}$\,=\,$|D(t$=$0)\rangle$ adiabatically follows $\ket{D(t)}$ without populating any excited states.
In the presence of strong interactions, however, population in the Rydberg state of one atom prevents another Rydberg excitation in its vicinity.
Thus, as one gradually turns on $g(t)$, any neighboring atoms initially in $\ket{++}$ necessarily populate the excited states, followed by their decay into one of the three ground states.
When $0$\,$<$\,$g$\,$\ll$\,$\Omega$, this dissipative dynamics produces effective jump operators of the form $c_\phi$\,=\,$\ket{\phi}\bra{DD}$ with a total rate
\begin{equation}
\Gamma_{DD}\approx-2\frac{g^4}{\Omega^4}\text{Im}\left[\frac{U}{1+i\chi U}\right],
\end{equation}
where $\chi$\,$\approx$\,$1/\gamma+ \gamma/(4\Omega^2)$, and $\phi$ is one of 9 different combinations of two-particle ground states (see Ref.\,\cite{EffOpOpenSys} and Appendix~\ref{sec:detail_REIT}).
To engineer the full Liouvillian, we can apply microwave pulses to the three ground states and globally rotate the spin-1 particles by $\theta$\,$\approx$\,$2\pi/5$ [Fig.\,\ref{fig:EITimpl}(b)].
When dissipative interactions and global rotations are alternated, this protocol effectively realizes a dynamics similar to $\mathcal{L}_\text{MP}$ and deterministically prepares AKLT states.
The experimental platform with a rearrangeable atom array in Refs.\,\cite{AtomArray,ParisAtomArray,AtomArrayPhysics,ParisAtomArrayPhysics,LesterAtomArray} is particularly well-suited to parallelize the implementation and exponentially shorten preparation times for large systems.

In practice, unwanted dissipations or interactions can affect the fidelity of our protocol by perturbing the steady state of dissipative dynamics.
There are two main imperfections in our proposed implementation: i) atomic states have finite dephasing time $T_2$, and ii) long-range Rydberg interaction can lead to dissipative coupling with particles beyond nearest neighbors.
For the latter, we find that pairs of particles separated by distance $R$ with interaction $U$\,$\sim$\,$1/R^6$ acquire decay rates $\Gamma_{DD}(R)$\,$\sim$\,$1/R^{12}$.
We study the effects of these imperfections by numerical simulations of long-range effective Hamiltonians and stochastic quantum jumps that now include dephasing operators $\ketbra{s}$ for $s$\,$=$\,$+,0,-$.
We introduce an effective temperature $T_\textrm{eff}$ defined by $\trace[P_{G} \rho(T_\textrm{eff})]$\,$=$\,$\F_\textrm{SS} $, where $\F_\textrm{SS}$ is the steady-state fidelity and $\rho(T_\textrm{eff})$\,=\,$\exp(-H_\textrm{AKLT}/T_\textrm{eff})/Z$ is the Gibbs ensemble with $Z$\,=\,$\textrm{tr}[\exp(-H_\textrm{AKLT}/T_\textrm{eff})]$ \cite{DeutschThermalization, EisertThermalization}.
When the steady state is near the gapped ground state, $T_\textrm{eff}$ characterizes the quality of prepared state in the thermodynamic limit.
The results in Fig.\,\ref{fig:EITimpl}(c) show that the temperature decreases with increasing dephasing time $T_2$ and eventually saturates due to long-range interactions.
While $T_\textrm{eff}$ also depends on system size $n$, we find that it stays below the gap of $H_\textrm{AKLT}$ for all $n$ studied in the present work (up to $n=8$).
We note that the effect of long-range interaction is mitigated in our parallelized protocol, where jump operators are turned on only for a few spin pairs well separated by the length of connected chains.
Also, throughout the parallelized protocol, the effective temperature does not increase, since our connection procedure ensures that $1-\F$ scales linearly in system sizes while the density of excited states grows at least as fast (see Appendix~\ref{sec:imperfection-scaling}).

\section{Generalization to MPS with symmetry\label{sec:generalization}}

Our symmetry-based approach can be generalized to efficiently prepare a broader class of matrix product states.
In general, any translation-invariant MPS of $n$ spins can be written as
\begin{equation}
\ket{A_{ab}^n} = \sum_{\{s_i\}} \braket{a|A^{(s_1)}A^{(s_2)}\cdots A^{(s_n)}|b}\ket{s_1s_2\cdots s_n},
\end{equation}
where $s_i\in\{1,2,\ldots,d\}$ runs over the physical spin basis for the $i$-th particle, and $\ket{a},\ket{b}\in\C^D$ indicate the ``boundary conditions'' in the virtual bond space.
For the case of AKLT states, we have $d=3$ and $D=2$.
Under periodic boundary condition, the unique MPS is given by $\ket{A_\circ^n} = \sum_a \ket{A_{aa}^n}$.
We say that such an MPS respects an internal symmetry group $\cG$ if for every $g\in \cG$ and some unitary representation $V_g: \cG\to U(d)$ we have
\begin{align}
V_g^{\otimes n} \ket{A_\circ^n} = e^{i\theta_g} \ket{A_\circ^n}.
\end{align}
Our protocol can be generalized to prepare any such translation-invariant MPS with internal symmetry, which is a class of states that includes ground states of all one-dimensional SPT phases\,\cite{MPSsymmetry,SchuchSPTMPS}.

\subsection{Generalized protocol with a minimal set of decay channels\label{sec:general-main}}
We now show how to generalize our protocol to any translation-invariant MPS with internal symmetry $\cG$.
Specifically, we design a dissipative dynamics that deterministically prepares the ground state(s) of the MPS parent Hamiltonian.
This protocol uses a set of global coherent manipulations corresponding to symmetry operations on the MPS, as well as a minimal number $k_\textrm{min}$ of decay channels $\{ c_1, c_2, \dots, c_{k_\textrm{min}} \}$ acting on pairs of neighboring particles.
We are able to derive a lower bound for $k_\textrm{min}$ based on irreducible representations of the symmetry group $\cG$, and provide an explicit construction of a minimal set of jump operators saturating the bound.
Given such a set $\{c_\mu\}$, the uniqueness of steady states can be efficiently verified via the same inductive proof technique in Sec.~\ref{sec:uniqunessAKLT}.
For the purpose of preparing ground states of SPT phases, we note that the symmetry $\cG$ of a parent Hamiltonian of the MPS may be larger than the minimal symmetry $\cG_p \subset\cG$ that protects the topological order.
For example, $\cG=\SO(3)$ while $\cG_p=\mathds{Z}_2\times \mathds{Z}_2$ in the case of AKLT states~\,\cite{HaldaneSPT}.

We now describe our results on the minimum complexity on the decay channels (i.e. jump operators) necessary to prepare a general translation-invariant MPS with symmetry.
Without loss of generality, we may assume that the desired states are ground states of a gapped, frustration-free
parent Hamiltonian $H_p =\sum_i h^{(i)}$, where $h^{(i)}$ is a translation-invariant, nearest-neighbor projector that respects the internal symmetry $\cG$~\cite{MPSreview,MPSsymmetry}.
Each term $h$ can be written in a block diagonal form, corresponding to different irreducible representations of $\cG$.
We refer to the two-particle subspace that $h$ projects onto as ``bright manifold'' $\B \equiv \range(h) \subset \mathds{C}^{d^2}$, where $d$ is the internal dimension of each particle.
The ground states are uniquely characterized by vanishing populations in $\B$ for every neighboring pair of particles.
In the $H_\AKLT$ example, $\B$ corresponds to the $J$\,=\,$2$ manifold of two neighboring spins.
Similar to our protocol for AKLT states, we can depopulate $\B$ by employing jump operators $c_\mu$ where $\range(c_\mu^\dag c_\mu)\subseteq \B$.
The number of jump operators can be reduced by utilizing and averaging over all symmetry rotations through $c_\mu \mapsto V_g^\dagger c_\mu V_g$, where $V_g$ is the global unitary rotation by a group element $g\in \cG$.
In order to fully depopulate the bright manifold $\B$ and nothing else, the set of jump operators $\{c_\mu\}_{\mu=1}^{k_{\min}}$ must satisfy the necessary condition
\begin{equation}
\label{eq:necessary-cond-for-jumps}
\B = \range\left(\sum_{\mu=1}^{k_{\min}} \sum_{g\in\cG} V_g^\dag c_\mu^\dag c_\mu V_g\right).
\end{equation}
As we show below, the minimum number $k_{\min}$ of distinct jump-operators will depends on the structure of the group representation of $\cG$.

For simplicity of discussion, let us restrict the decay channels to the rank-1 form of $c_\mu = \ketbrat{\phi_\mu}{\psi_\mu}$.
The minimum number of jump operators required can be calculated from the number of different irreducible representations (irrep) of the symmetry group $\cG$ within the bright manifold $\B$.
In the case of AKLT states, $\B$ consists of a single 5-dimensional irrep of the group $SO(3)$.
In more general cases, the representation of $\cG$ on $\B$ may contain multiple copies of isomorphic (i.e., equivalent up to a basis change) irreps.
The capability of global symmetry operations allows one decay channel to depopulate subspaces corresponding to one copy of each irrep in parallel.
Hence, as we have shown earlier, one decay channel is sufficient for preparing the AKLT state.
For the more general cases, however, it may be necessary to employ multiple decay channels when more than one copy of an irrep is present.
Using Schur's lemma~\cite{ArtinAlgebra}, we prove that the minimum number of rank-1 decay channels is
\begin{equation}
k_{\min} = \max_{\substack{\text{irrep } r \text{ of } \cG \text{ in } \B}} \left\lceil \frac{\text{\# of copies of irrep } r  \text{ in }\B }{\text{dimension of irrep }r}\right\rceil.
\end{equation}
We can also construct $k_{\min}$ decay channels that satisfy the necessary condition of Eq.~\eqref{eq:necessary-cond-for-jumps} by choosing a set of $\{\bra{\psi_\mu}\}$ that is supported in all irreps, with destructive interference between isomorphic irreps.
The details of the proof and the construction are described in Appendix~\ref{sec:general}.

Once we have such a minimal set of decay channels $\{c_\mu\}$, it remains to ascertain the uniqueness of steady states.
This can be efficiently verified using our inductive proof techniques, which show that the steady states are unique as long as there are only trivial solutions to a linear equation like \eqref{eq:uniqueness}.
More specifically, to prove uniqueness under open boundary condition, one simply needs to compute a $D^2 \ell \times D^2 \rank(\B)$ matrix $\vect{M}$ whose matrix elements are
\begin{align}
(\vect{M})^{abz}_{pq\nu}=
\sum_{r=1}^D\braket{pa|\T^{n-2}|rb} \times\braket{G_{rq}^2|c_\nu|z},
\end{align}
where $\T$ is the transfer matrix for the MPS, $a,b,p,q,r,c$ are indices for $D$-dimensional virtual bond space, and $\ket{z}$ enumerates the possible states in the bright manifold $\B$.
Here, $\ell \le d^2$ is related to the maximum dimension of the irreps of the symmetry group $\cG$ on tthe bright manifold $\B$. 
By verifying (through exact calculations or numerics) that the steady state is unique for some small system size $n_0$, one can then prove uniqueness for all $n\ge n_0$ by showing that $\det(\vect{M}^\dag \vect{M})\neq 0$.
We emphasize that these proofs can be done efficiently for any given $\{A^{(s)}\}$ and decay channels $\{c_\mu\}$, since the calculations only involve matrices of constant dimensions, independent of system size $n$.

Our Rydberg-EIT implementation proposal can be naturally adapted for these general cases.
The Rydberg-EIT scheme allows us to engineer two-body jump operators of the form $c_\eff=\ketbrat{s^{L1}s^{L2}}{s^Rs^R}$, where $\ket{s^{Li}}, \ket{s^R}\in\C^d$ are single-spin states.
Unlike the case of AKLT states, the preparation of a generic symmetric MPS may require more than one ($k_\textrm{min}\ge2$) rank-1 jump operator.
The implementation of multiple decay channels can be achieved, for example, by introducing extra lasers that couple (additional) hyperfine ground states to the short-lived excited state(s).
By adjusting the relative strength of laser driving to each hyperfine ground state, one can engineer different EIT-dark states that acquire a dissipative interaction.
This allows us to generate a set of jump operators $\{c^\eff_\mu = \ketbrat{s^{L1}_\mu s^{L2}_\mu}{s^R_\mu s^R_\mu}: \mu = 1,...,k_\text{min}\}$ with independent $\ket{s^R_\mu s^R_\mu}$.
When particles are individually addressable, one can engineer a jump operator with a more complicated right-singular vector $\bra{\psi_\mu}$, i.e. $\bra{\psi_\mu}\neq \bra{s^R_\mu}^{\otimes2}$ for any $\bra{s^R_\mu}\in \C^d$.
For example, if we can engineer a unitary $U$ where $\bra{s^R_\mu}^{\otimes 2} U = \bra{\psi_\mu}$ and $U^\dag\ket{s_\mu^{L1}s_\mu^{L2}}=\ket{\phi_\mu}$, then applying $U$ stroboscopically each time before turning on dissipative interaction [see Fig.~\ref{fig:EITimpl}(b)] would allow engineering of $c_\mu = \ketbrat{\phi_\mu}{\psi_\mu}$.

\subsection{Generalizing the parallelized protocol\label{sec:general-parallel}}
We can also extend our strategy of parallelized connection and quantum feedback to the class of translation-invariant MPSs with internal symmetry.
Recall that the idea is to prepare many segments of the desired MPS with open boundary conditions, and then connecting adjacent pairs of segments in parallel.
The analysis is much simpler when the MPS is \emph{injective}, which means that largest-magnitude eigenvalue of the transfer matrix $\T=\sum_{s} A^{(s)*}\otimes A^{(s)}$ is non-degenerate.
When the desired MPS is injective, which is true for generic cases\,\cite{MPSreview} (and also for AKLT states), we can show that the success probability of connection is typically at least $1/D^2$.
This system-size-independent success probability implies that the scaling of preparation time of the parallelized protocol is $\O(\log^2 n)$, which exponentially outperforms existing dissipative protocols that do not involve parallelization and feedback\,\cite{Verstraete}.
The case of non-injective MPSs is more subtle, since the associated parent Hamiltonian of the MPS has degenerate ground states.
These states can be shown to be the only steady states of our protocol, but typically a mixture of them will be prepared.
Nevertheless, if we place some (often reasonable) restrictions on the initial state, a pure non-injective MPS can be prepared, as we illustrate with the example of a Greenberger-Horne-Zeilinger (GHZ) state.

\textbf{Injective case}---
We first analyze the protocol for the case of injective MPSs.
Consider an arbitrary initial state of two adjacent length-$m$ chains of MPSs, which can be written as $\ket{\psi_0} = \N_0\sum_{b,c} C_{bc} \ket{A_{ab}^{m}} \otimes\ket{A_{cd}^{m}}$, where $C_{bc}\in\C^{D\times D}$ is some coefficient matrix that characterizes the edge states at the interface of the chains, and $\N_0$ is a normalization constant.
If the two chains are unentangled, then $C_{bc} = \alpha_b\beta_c$ for some $\vec\alpha,\vec\beta\in\C^D$.
By turning on the jump operators acting at the interface, we can cool this state into the desired final state $\ket{\psi_f^{2m}} = \N_f \ket{A_{ad}^{2m}}$, for some normalization constant $\N_f$.
Since the MPS is assumed to be injective, the success probability of connection can be shown to be
\begin{align}
p &= \left|\braket{\psi_f^{2m}|\psi_0}\right|^2 = \frac{\left|\tr(C)\right|^2}{D\tr(C^\dag C)} + \O(\epsilon_2^m) \nonumber \\
&= \frac{|\vec\alpha\cdot\vec\beta|^2}{D|\vec\alpha|^2|\vec\beta|^2} + \O(\epsilon_2^m) \quad \text{if} \quad C_{bc} = \alpha_b\beta_c,
\end{align}
where $\epsilon_2$ is the second largest eigenvalue of $\T$ (see Appendix~\ref{appx:generalization} for more details). 
For random states $\vec\alpha,\vec\beta\in\C^D$, we have on average $p\simeq1/D^2$.
The maximum success probability of $p_\text{max}=1/D$ is obtained when $\vec\alpha \parallel \vec\beta$, i.e., when the two edge states are identical.

When our desired MPS exhibits bond-inversion symmetry $\mathcal{P}$, as the AKLT states do, we have the same issue of vanishing success probability after a quantum jump that also respects $\mathcal{P}$.
To see this, consider what happens in the event of a quantum jump due to a jump operator of the form $c=\ketbrat{\phi}{\psi}$.
The state after discarding the two particles at the interface is $\ket{\psi_1} =\N_1\sum_{b',c'}\tilde{C}_{b'c'}\ket{A_{ab'}^{m-1}}\ket{A_{c'd}^{m-1}}$, where
\begin{equation}
\tilde{C}_{b'c'} = \sum_{bc}\bra{\psi}(\ket{A_{b'b}^1}\otimes\ket{A_{cc'}^1}) C_{bc}.
\end{equation}
If $\ket{\psi}$ respects bond-inversion symmetry, i.e. $\mathcal{P}\ket{\psi}=\pm\ket{\psi}$, where $\mathcal{P}=\sum_{i,j}\ketbrat{ij}{ji}$ is the swap operator, then
\begin{align}
\tr(\tilde{C}) &= \sum_{a} \tilde{C}_{aa}
= \sum_{abc} \bra{\psi}(\ket{A_{ab}^1}\otimes\ket{A_{ca}^1}) C_{bc} \nonumber\\
&= \sum_{abc} \bra{\psi}\mathcal{P} (\ket{A_{ca}^1}\otimes\ket{A_{ab}^1}) C_{bc} \nonumber \\
&=
\sum_{bc}\pm \braket{\psi|A_{cb}^2}C_{bc} = 0.
\end{align}
We find that this quantity is zero regardless of the initial $C_{bc}$, entangled or unentangled, due to our requirement that $\ket{\psi}$ be orthogonal to the desired MPS $\ket{A_{cb}^2}$.
In fact, $\tr(\tilde{C})$ is related to the success probability of the next connection attempt, $p'=\big|\braket{\psi_f^{2m-2}|\psi_1}\big|^2 \propto |\tr(\tilde{C})|^2+\O(\epsilon_2^m) = \O(\epsilon_2^m)$, which is exponentially small for a large system size $m$.

Similar to the AKLT case, we can also try to restore the success probability by applying a global symmetry operation $U_g^{\otimes m-1}$ for some $g\in\cG$ to one of the chains.
In our AKLT protocol, there is a symmetry operation $U_g=e^{i\pi S_y}$ whose action on the virtual bond level $u_g=e^{-i\pi\sigma_y/2}$ yields $\big|\tr(u_g^\dag \tilde{C})\big|^2=\tr(\tilde{C}^\dag\tilde{C})$, allowing us to recover the maximum success probability of $p_\text{max}=1/D$ regardless of the initial state $C_{bc}$ or which quantum jump occurred.
While the existence of such an operation is not known for the general case, we can at least restore the success probability to $1/D^2$ for many injective MPSs by applying a randomly chosen symmetry operation.
This is because injectivity is typically associated with the irreducibility of the representation $u_g$ on the virtual bond level~\cite{MPSsymmetry}, which allows us to show that $p= 1/D^2$.
If we write $\tilde{C} = \sum_i \sqrt{\lambda_i} \ketbrat{b_i}{a_i}$ in its singular value decomposition form, then 
\begin{align}
\left|\tr(u_g^\dag \tilde{C})\right|^2 &= \sum_{i,j}\sqrt{\lambda_i \lambda_j} \braket{a_i|u_g^\dag|b_i} \braket{b_j|u_g|a_j}.
\end{align}
If $u_g$ is irreducible, then by Schur's lemma~\cite{ArtinAlgebra} we have
\begin{align}
\mathds{E}_{g\in \cG}\left[\big|\tr(u_g^\dag \tilde{C})\big|^2 \right] &=  \sum_{i,j}\sqrt{\lambda_i \lambda_j} \braket{a_i|a_j}\braket{b_j|b_i}/D \nonumber \\
&= \frac{1}{D}\sum_{i} \lambda_i = \frac{1}{D} \tr(\tilde{C}^\dag \tilde{C}).
\end{align}
Hence, this yields a subsequent success probability of $p\simeq1/D^2$.
A sufficient condition for the irreducibility of $u_g$ is that $U_g$ be irreducible and $\{A^{(s)\dag} A^{(s')}:\forall s,s' \}$ spans the whole space of $D\times D$ matrices (see Proposition 17 in Ref.~\cite{MPSsymmetry}).

\textbf{Non-injective case}---
For a non-injective MPS, the analysis is complicated by the presence of degenerate ground states of its parent Hamiltonian.
Here we consider an illustrative example to prepare GHZ states: $\ket{\GHZ_\pm} = (\ket{0^n}\pm\ket{1^n})/\sqrt{2}$.
These states have an MPS representation with $(d,D)=(2,2)$ given by the following matrices:
\begin{equation}
\begin{split}
A^{(0)}=\ketbra{\up} &= \begin{pmatrix} 1 & 0 \\ 0 & 0 \\ \end{pmatrix}\\
\text{and} \quad
A^{(1)}=\ketbra{\dn} &= \begin{pmatrix} 0 & 0 \\ 0 & 1 \\ \end{pmatrix}.
\end{split}
\end{equation}
In this representation, $\ket{\GHZ_+} \propto \ket{A^n_{\rightarrow\rightarrow}}=\ket{A^n_{\leftarrow\leftarrow}}$, and $\ket{\GHZ_-} \propto \ket{A^n_{\rightarrow\leftarrow}} = \ket{A^n_{\leftarrow\rightarrow}} $, where $\ket{\rightarrow} = (\ket{\up}+\ket{\dn})/\sqrt{2}$ and $\ket{\leftarrow} = (\ket{\up}-\ket{\dn})/\sqrt{2}$ are possible edge configurations.
This MPS has an internal symmetry group of $\cG=\mathbb{Z}_2$, which is represented by $\{\I,\sigma_x^{\otimes n}\}$ acting on the system.
Its parent Hamiltonian is $H_\GHZ = \sum_{i} (\I -\sigma_z^{(i)}\sigma_z^{(i+1)})$, whose ground states are doubly degenerate due to non-injectivity.
The corresponding two-particle bright manifold is $\B=\text{span}\{\ket{\Phi_+},\ket{\Phi_-}\}$, where $\ket{\Phi_\pm} = (   \ket{01}\pm\ket{10})/\sqrt{2}$.
The two states $\ket{\Phi_\pm}$ support two distinct irreducible representations of $\mathds{Z}_2$, which are the trivial and the sign representation, respectively.
Hence, we can use just one jump operator of the form e.g. $c=\ket{00}(\kappa_+\bra{\Phi_+} + \kappa_-\bra{\Phi_-})$, with $\kappa_\pm\neq 0$ so that both irreps are supported (a necessary condition as shown in Appendix~\ref{sec:general}).
Then along with the global symmetry operation $\sigma_x^{\otimes n}$, we can depopulate the bright manifold and obtain $\text{span}\{\ket{\GHZ_\pm}\}$ as the subspace of steady states.

Now let us consider preparing $\ket{\GHZ_\pm}$ in a parallelized protocol with connections and feedback.
We note that unlike in the injective case, different choices of jump operator here can lead to qualitatively different outcomes.
Specifically, we consider two choices of jump operators that may result in different degrees of entanglement of the final state.
First, consider an example choice of jump operator $c=\ketbrat{00}{01}$ (i.e. $\kappa_\pm=1/\sqrt{2}$).
While this along with the symmetry operation produces a dissipative dynamics that has $\ket{\GHZ_\pm}$ as the steady states, the parallelized protocol can only produce an unentangled final state of either $\ket{0^n}\propto\ket{\GHZ_+}+\ket{\GHZ_-}$ or $\ket{1^n}\propto\ket{\GHZ_+}-\ket{\GHZ_-}$ once any quantum jump occurs, regardless of states of the initial chains.
Alternatively, we may choose the jump operator $c=\ket{00}(\bra{01}+i\bra{10})/\sqrt{2}$ (i.e., $\kappa_+=\kappa_-^*=(1+i)/2$).
In this case, suppose we start with $\ket{0^{n_0}}+\ket{1^{n_0}}$ on the initial chains of length $n_0$, then we can produce a maximally entangled final state of $\ket{0^n}+ \zeta\ket{1^n}$ even after quantum jumps, for some $\zeta\in\{\pm1,\pm i\}$ that we can determine from recording quantum jump history.
In both cases, for an arbitrary (unentangled) initial state $\ket{\psi_0}=\sum_{a,b=0}^1(\alpha_a\ket{a^m})\otimes (\beta_b\ket{b^m})$ of two chains, the success probability of connecting them is on average $1/2$ for random $\vec{\alpha},\vec{\beta}\in\C^2$.
This system-size-independent success probability means that the parallelized protocol for this non-injective MPS also has an efficient scaling of $\O(\log^2 n)$ for the preparation time.

\section{Summary and Outlook\label{sec:outlook}}
In this work, we have shown that using only one type of nearest-neighbor decay channel and global control, an AKLT state in a large system can be efficiently prepared as the provably unique steady state of the driven-dissipative dynamics.
Since the symmetry group of AKLT states is continuous, the decay channels and pulses need not be fine-tuned, and the implementation is robust against imperfections in experimental parameters.
The scaling of preparation time was numerically shown to be polynomial, and can be improved exponentially using parallelization and quantum feedback.
This proposal is feasible for a wide range of controlled quantum systems, and in particular optimal for a rearrangeable array of trapped neutral atoms using a Rydberg-EIT scheme.
In addition, we analyzed the generalization to other translation-invariant MPSs with symmetry, and derived a bound on the minimum number of necessary decay channels that may be saturated by a construction.
We also showed that the parallelized protocol can work in the general case, and provided some sufficient conditions for success, such as the injectivity of the MPS and the irreducibility of the symmetry group representation on the virtual bond level.

Finally, we note that it may be possible to generalize our symmetry-based dissipative preparation scheme to higher dimensional many-body entangled states.
Many such states are described by projected entangled pair states (PEPS), a natural generalization of MPSs for arbitrary lattices, which also allow construction of frustration-free parent Hamiltonians\,\cite{PEPSparent} to be converted into jump operators\,\cite{Verstraete,KrausZoller,TicozziViola2012}.
However, our inductive proof of uniqueness of steady states does not extend straightforwardly, since exact computation of expectation values of a generic PEPS is intractable\,\cite{PEPScomplexity}.
Further investigations are thus necessary to extend our strategy to higher dimensions, which can support even more interesting, long-range entangled states with symmetry-enriched topological order~\cite{MesarosSET}.

\begin{acknowledgments}
We thank H. Pichler and F. Reiter for useful discussion. We thank Haoyang Gao for bringing to our attention a technical problem in a previous version of the manuscript. This work was supported through NSF, CUA, Vannevar Bush Faculty Fellowship, AFOSR Muri, and Moore Foundation. L.Z. was supported by NSF Graduate Research Fellowship under Grant No.~DGE1144152. S.C. was supported by Kwanjeong Educational Foundation.
\end{acknowledgments}

\appendix

\section{Effective Liouvillian\label{sec:eff-liouvillian}}
Our proposal for preparing AKLT states uses only one type of jump operator, e.g. $c=\ketbrat{00}{++}$.
The dissipative dynamics due to this jump operator is $\L_0= \Gamma \sum_i \D[c^{(i)}] $, where $i$ enumerates the pair of neighboring sites $(i,i+1)$. Our key idea is to use coherent global manipulations, corresponding to operations in the symmetry group, so as to effectively realize additional jump operators.
This can be achieved either by periodically applying pulsed global spin rotations (symmetry operations), or by continuously rotating the spins.

In the first, multi-pulse sequence approach, we apply pulses $V_\theta = (e^{i\theta S_y})^{\otimes n}$, each separated by an interval of $\tau$. Then in the rotating frame, we have the time-dependent Liouvillian:
\begin{equation}
\L_\MP(t)= (V_\theta^\dag)^k \L_0 V_\theta^k \quad \text{for} \quad  k\tau \le t < (k+1)\tau.
\end{equation}
Suppose we choose $\theta=2\pi/\ell$ for some integer $\ell$, then this dynamics is periodic with period $\ell\tau$ , since $V_{2\pi/\ell}^\ell=\I$. In the limit of fast pulses $\tau\ll1/\Gamma$, we can use the first-order Magnus expansion to derive a simpler, effective time-independent Liouvillian that approximate the dynamics:
\begin{equation}
\bar{\L}_\MP = \frac{1}{\ell\tau}\int_0^{\ell\tau} \L_\MP(t) dt = \frac{\Gamma}{\ell}\sum_i \sum_{\nu=0}^{\ell-1} \D[\bar{c}_\nu^{(i)}],
\end{equation}
where $\bar{c}_\nu = (V_{2\pi/\ell}^\dag)^\nu \, c \, V_{2\pi/\ell}^\nu$.

Alternatively, we may employ a continuous wave approach by introducing a time-independent Hamiltonian $H_\CW=\omega\sum_iS_i^y$. Then in the rotating frame, we have $c(t)=e^{i\omega t S_y}c e^{-i\omega tS_y}$, and
\begin{equation} \label{eq:CWLiou}
\dot\rho = \L_\CW(t)\rho \equiv \Gamma\sum_{i}\D[c^{(i)}(t)]\rho.
\end{equation}
In this frame, the dynamics is periodic with period $2\pi/\omega$.
Again, we compute the effective time-independent Liouvillian
\begin{equation}
\bar{\L}_\CW = \frac{\omega}{2\pi}\int_0^{2\pi/\omega} \L_\CW(t)dt = \sum_{i}\sum_{\beta=0}^{\ell-1} \Gamma_\beta \D[\bar{c}_{\beta}^{(i)}],
\end{equation}
obtained by time-averaging (first-order Magnus expansion) as an approximation.
The effective jump operator $\bar{c}_{\beta}$ in the standard form of a Liouvillian can be obtained by diagonalizing the superoperator acting on the space of density operators. More explicitly, we diagonalize a Hermitian matrix $L=U^\dag \Lambda U$, whose entries $L_{ii',jj'}$ are given by
\begin{align}
L_{ii',jj'} &\equiv \Braket{ij|\frac{\omega}{2\pi}\int_0^{2\pi/\omega} dt~ \Gamma c^*(t)\otimes c(t)|i'j'} \nonumber \\
&= \sum_\beta U_{\beta,ii'}^* \Lambda_\beta  U_{\beta,jj'} \equiv \Braket{ij|\sum_\beta \Gamma_\beta \bar{c}_\beta^*\otimes \bar{c}_\beta|i'j'}.
\end{align}
We can read off $\Gamma_\beta=\Lambda_\beta$, and $\bar{c}_\beta=\sum_{k,k'}U_{\beta,kk'}\ketbrat{k}{k'}$.
For our example choice of $c(t=0)=\ket{00}\bra{++}$, we obtain $\ell=9$ independent jump operator after evaluating the integral and diagonalizing, each with rate $\Gamma_\beta/\Gamma = 7/32$, 3/16, 3/16, 1/8, 1/8, 1/16, 1/16, 1/64, and 1/64, respectively.

\section{Quantum evolution conditioned on no quantum jump\label{sec:quantum_jump}}

To understand how our quantum state evolves conditioned on detecting no quantum jumps, we use the stochastic wavefunction formalism for open system dynamics.
Namely, we define an effective non-Hermitian Hamiltonian
$H_\eff = H -  i \sum_\mu\Gamma_\mu c_\mu^\dag c_\mu/2$, where $H$ is the Hamiltonian of the system and $c^{(i)}$ are quantum jump operators. A system evolves under $H_\textrm{eff}$ until it undergoes a quantum jump $\ket{\psi} \mapsto c_\mu \ket{\psi}$ at a rate $\bra{\psi}\Gamma_\mu c_\mu^\dag c_\mu \ket{\psi}$.
In our protocol to prepare AKLT states, $H = 0$ since we work in the rotating frame.
Hence, $H_\eff$ is anti-Hermitian and thus diagonalizable with eigenvalues $\lambda_\alpha =  -i\gamma_\alpha/2$.
It is assumed that our desired states are ``dark states'', which are eigenvectors of $H_\eff$ with a zero imaginary part of the eigenvalue.
For simplicity and illustrative purposes, let us also assume there is just one dark state $\ket{0}$ with eigenvalue $\lambda_0=0$, and the rest of the eigenvalues are sorted by $0<\gamma_1\le \gamma_2\le \cdots$.
Let us decompose the initial state in the eigenbasis $\ket{\psi_0} = \sum_\alpha c_\alpha \ket{\alpha} = c_0\ket{0} + c_1\ket{1}+\cdots$.
The evolution under $H_\eff$ yields the unnormalized state
\begin{equation}
\ket{\tilde\psi(t)} \equiv e^{-iH_\eff t}\ket{\psi_0} = c_0 \ket{0} + c_1e^{-\gamma_1t/2}\ket{1}+\cdots.
\end{equation}
The probability of undergoing no quantum jump over a time duration $T$ is
\begin{align}
p_0(T) &= \braket{\tilde{\psi}(T)|\tilde{\psi}(T)} = |c_0|^2 + |c_1|^2e^{-\gamma_1T} + \cdots \nonumber \\
& \ge |c_0|^2 = \left|\braket{0|\psi_0}\right|^2.
\end{align}
Conditioned on such an event, the fidelity of the quantum state preparation is
\begin{align}
\mathcal{F}(T) &=  \frac{\left|\braket{0|\tilde{\psi}(T)}\right|^2}{\braket{\tilde{\psi}(T)|\tilde{\psi}(T)}} = \frac{|c_0|^2}{|c_0|^2+|c_1|^2e^{-\gamma_1T}+\cdots}  \nonumber \\
&=  1 - \O(e^{-\gamma_1T}) \quad \text{ if } \quad |c_0|^2>0,
\end{align}
where we find that the fidelity exponentially approaches unity.
This allows for effective ``cooling'' of the system into the desired state when there is no quantum jump for $T\gg1/\gamma_1$, which occurs with probability $p_0\simeq \left|\braket{0|\psi_0}\right|^2$.
It is also easy to see that when there are multiple dark states, given by some projector $P_D$, the system is cooled into $\ket{\psi_0}\to P_D\ket{\psi_0}$, conditioned on no quantum jumps which occurs with probability $p_0 = \braket{\psi_0|P_D|\psi_0}$.

\section{Proof of uniqueness of steady states for the AKLT example\label{sec:proof_uniqueness}}
In this appendix, we provide a detailed proof of our lemma that $\bar{\L}_\MP$ and $\bar{\L}_\CW$ have AKLT states as unique steady states, for $n\ge 2$ under the open boundary condition.
Although the proof has already been sketched in the main text, here we provide a more detailed analysis.
Let us define $\L_i=\sum_{\nu=0}^{\ell-1} \Gamma_\nu \D[\bar{c}_{\nu}^{(i)}]$ as the Liouvillian acting only on sites $i$ and $i+1$, and $\L_{[n]} = \sum_{i=1}^{n-1}\L_i$.
Let $Q_{i,i+1}$ be the projector onto the $J=2$ manifold on spin $i$ and $i+1$.
We also define 
\begin{align}
H_\AKLT^{[n]} = \sum_{i=1}^{n-1} Q_{i,i+1}
\end{align}
as the AKLT Hamiltonian acting on $n$ spins under the open boundary condition.
We also let $D_n = \text{span}\{\ket{G_{ab}^{n}}\}$ be the space of AKLT states on $n$ spins, and let $P_n$ be its corresponding projector.
We prove by induction on $n$.

We start with our inductive hypothesis, which is that $\L_{[n]}$ only has the AKLT states $\ket{G_{ab}^{n}}$ as steady states.
This means that any $n$-particle wavefunction can reach $D_n$ via some sequence of jump operators $\bar{c}_\nu$.
More formally, this means that for any $n$-particle wavefunction $\ket{\psi^n}$, there exists a polynomial function of jump operators $f_{[n]}(\{\bar{c}_\nu\})$ such 
that 
\begin{align}
\| P_n f_{[n]} \ket{\psi^n} \| \neq 0 \,.
\end{align}
We only need to show that any $(n+1)$-particle wavefunction $\ket{\psi^{n+1}}$ can reach $D_{n+1}$ via jump operators, which would yield the proof of our lemma.

From this inductive hypothesis, we first argue that there exists a polynomial $f_{[n]}(\{\bar{c}_\nu\})$ of jump operators acting on the first $n$ spins such that 
\begin{align} \label{eq:popu-n-nonzero}
\|P_n f_{[n]}\ket{\psi^{n+1}}\| \neq 0 \,.
\end{align}
This can be done by simply looking at the the reduced density matrix $\rho_n$ of $\ket{\psi^{n+1}}$ on the first $n$ spins, and picking $f_{[n]}$ with respect to one of the non-trivial eigenvectors of $\rho_n$.
We note that the AKLT Hamiltonian is frustration free, so any (ground) state minimizing the energy of $H^{[n]}_\textrm{AKLT}$ must also minimize the energy of $H^{[n-1]}_\textrm{AKLT}$, or $P_{n} P_{n-1} = P_n$.
Hence \eqref{eq:popu-n-nonzero} implies $f_{[n]}\ket{\psi^{n+1}}$ also contains nonzero population in the AKLT ground states among the first $n-1$ spins, i.e.,
$\|P_{n-1} f_{[n]}\ket{\psi^{n+1}}\| \neq 0$.

Let us then perform a general decomposition \
\begin{equation}
f_{[n]}\ket{\psi^{n+1}} = \ket{\phi} + \ket{\phi^\perp_1} + \ket{\phi^\perp_2}
\end{equation}
where
\begin{align}
\ket{\phi} = \sum_{a,b=\dn}^\up \sum_{z=-2}^2 \phi_{abz}\ket{G_{ab}^{n-1}} \ket{z}.
\end{align}
Here $\ket{z}$ runs through the 5 states in the $J=2$ manifold on the last two spins.
The remaining parts of the wavefunction $f_{[n]}\ket{\psi^{n+1}}$ are
\begin{align}
\ket{\phi^\perp_1} &= \sum_{a,b,s,t} \phi_{abst}^\perp \ket{G_{ab}^{n-1}}\ket{G_{st}^2}, \\
\ket{\phi^\perp_2} &=\sum_{\mu, s} \phi_{\mu s}^\perp \ket{E_\mu^{n-1}}\ket{s},
\end{align}
where $\ket{E_\mu^{n-1}}$ runs through all the excited eigenstate of the $H^{[n-1]}_\AKLT$, and $s$ runs through all 9 possible 2-spin states.

We assume for the sake of contradiction that $\ket{\psi_{n+1}}$ is a state that cannot reach $D_{n+1}$ via jump operators.
We now consider two cases:

\textbf{Case 1:} $\phi_{abz}$\,=\,0 for all $a,b,z$. ---
Then $f_{[n]}\ket{\psi^{n+1}} = \ket{\phi^\perp_1} + \ket{\phi^\perp_2}$.
From our inductive hypothesis, $f_{[n]}\ket{\psi^{n+1}}$ must have nonzero population in $D_n$ and $D_{n-1}$, and this can only come from the $\ket{\phi^\perp_1}$ part, and thus $\phi_{abst}^\perp$ cannot be all vanishing.
On the other hand, since we have assumed $\ket{\psi^{n+1}}$ cannot reach $D_{n+1}$, we must have
$0= \braket{G_{pq}^{n+1}|\phi^\perp_1}$.
That means $Q_{i,i+1} \ket{\phi^\perp_1} \neq 0$ for some $i$.
Since $Q_{i,i+1}\ket{\phi^\perp_1} = 0$ for all $i\neq n-1$, this means that we must have  $Q_{n-1,n}\ket{\phi^\perp_1} = \ket{\phi^\perp_1}$.
Then
we have $P_n f_{[n]}\ket{\psi^{n+1}} = P_n \ket{\phi^\perp_1} + P_n\ket{\phi^\perp_2} = 0$,
since $P_nQ_{n-1,n} = 0$ and $P_n\ket{E_\mu^{n-1}}=0$. This gives a contradiction with \eqref{eq:popu-n-nonzero}.

\textbf{Case 2:} Some $\phi_{abz}$\,$\neq$\,$0$. ---
Consider any jump operator $\bar{c}_\nu^{(n)}$ acting on the spins $n$ and $n+1$. Note that
\begin{align}
\braket{G_{pq}^{n+1}| \bar{c}_\nu^{(n)} | \phi^\perp_2}& =\sum_{r,\mu, s} \braket{G_{pr}^{n-1}| E_\mu^{n-1}} \braket{G_{rq}^2| \bar{c}_\nu^{(n)} | s} \phi^\perp_{\mu s} \nonumber \\
&=0
\end{align}
because $\ket{G_{pr}^{n-1}}$ are ground states of $H^{[n]}_\AKLT$ and thus orthogonal to $\ket{E_\mu^{n-1}}$.
We also have $\braket{G_{pq}^{n+1}| \bar{c}_\nu^{(n)} | \phi^\perp_1} = 0$ since $\bar{c}_\nu\ket{G_{st}^2} = 0$.
Then our assumption that $\ket{\psi^{n+1}}$ cannot reach $D_{n+1}$ gives the condition
\begin{align}
0 &=\braket{G_{pq}^{n+1}| \bar{c}_\nu^{(n)} f_{[n]} | \psi^{n+1}}    \nonumber \\
&= \sum_{a,b=\dn}^\up \sum_{z=-2}^2 \braket{G_{pq}^{n+1}| \bar{c}_\nu^{(n)}  | G^{n-1}_{ab}} \ket{z} \phi_{abz} \nonumber \\
&= \sum_{a,b=\dn}^\up  \sum_{z=-2}^2 M_{pq\nu}^{abz} \phi_{abz}
\equiv \vect{M}\vec{\phi},
\label{eq:open-uniqueness}
\end{align}
where $\vec{\phi}$ is a $2\times 2\times 5=20$-dimensional vector, and $\vect{M}$ is a $4\ell\times$20 dimensional matrix.
The matrix elements of $\vect{M}$ are given by
\begin{align}
M_{pq\nu}^{abz} &=\braket{G_{pq}^{n+1}| \bar{c}_\nu^{(n)}  | G^{n-1}_{ab}} \ket{z}\nonumber \\
&= \sum_{r=\dn}^\up \braket{G_{pr}^{n-2}|G_{ab}^{n-1}} \braket{G_{rq}^2| \bar{c}_\nu | z}
\end{align}
and can be calculated analytically, since the first factor comes from diagonalizing the transfer matrix, and the second factor is computed in a small, 9-dimensional Hilbert space of two spins.
Now if $\det(\vect{M}^\dag \vect{M}) \neq 0$, then the matrix $\vect{M}$ has full rank, indicating that we only have the trivial solution $\phi_{abz}=0$.
This would contradict this case's assumption, and hence any $\ket{\psi_{n+1}}$ must be able to reach $D_{n+1}$ via jump operators, and the AKLT states $\ket{G_{ab}^{n+1}}$ are the unique steady states of $\L_{[n+1]}$.

Therefore, it only remains to compute $\det(\vect{M}^\dag \vect{M})$ and show that it is nonzero for $\L$ in our proposals.
Let us first consider $\bar{\L}_\MP$, with $\ell=5$ corresponding to rotation pulses with $\theta=2\pi/5$. For this, we explicitly find
\begin{eqnarray}
\det(\vect{M}^\dag \vect{M}) = \frac{5^{20} (x+3)^{30} (x-9)^{10} } {2^{100} 3^{36}x^{40}}
\end{eqnarray}
where $x= (-3)^n$.
It's easy to see that this is only zero for $n=1,2$, so our inductive proof holds for $n\ge 3$. The base case of $n=2, 3$ can easily be checked numerically or by exact calculations.

For $\bar{\L}_\CW$, which has $\ell=9$, we find
\begin{eqnarray}
\det(\vect{M}^\dag \vect{M})  
= \frac{(x+3)^{30} (x-9)^{10}}{2^{24} 3^{44} 7^4 x^{40}}
\end{eqnarray}
where again we have defined $x= (-3)^n$. This is also only nonzero when $n=1,2$.
Since the base cases of $n=2,3$ can be checked exactly, this proves that the steady states of $\bar{\L}_\MP$ and $\bar{\L}_\CW$ are uniquely given by $D_n$ for $n\ge2$.

Our proof method here naturally suggests a method to prepare an AKLT state with specified edge states instead of a mixture of the four $\ket{G_{ab}^n}$ under open boundary condition.
For instance, $\ket{G_{\up \up}^n}$ can be deterministically prepared by adding two jump operators: $c_L$\,=\,$\ketbrat{0}{-}_1$ on the left edge and $c_R$\,=\,$\ketbrat{0}{+}_n$ on the right.
In this case, it is easy to see that any linear combination of four ground states $\ket{G_{ab}^n}$ can further decay into $\ket{G_{\up\up}^n}$, which becomes the unique steady state.

\section{Details of our numerical simulation \label{sec:detail_numerics}}
We simulate the dissipative dynamics $\L=\sum_\mu \Gamma_\mu \D[c_\mu]$ using the stochastic wavefunction method~\cite{gardiner2004quantum}.
In this approach, the wavefunction $\ket{\psi(t)}$ continuously evolves under the effective non-Hermitian Hamiltonian $H_\eff = H-i\sum_\mu \Gamma_\mu c_\mu^\dag c_\mu/2$ and stochastically undergoes quantum jumps $c_\mu$ at a rate $\braket{\psi|\Gamma_\mu c_\mu^\dag c_\mu|\psi}$.
Physical observables are extracted from an ensemble of wavefunction trajectories obtained from independent simulations.
Compared to direct numerical integrations of quantum master equations, this method allows simulation of systems with a larger number of particles.
In order to simulate a maximally mixed initial state (or equivalently an infinite temperature ensemble), we sample a random product state in the $S_z$-basis as the initial state $\ket{\psi_\alpha(0)}$ for the $\alpha$-th simulation.
For each small time step $\delta t$, the state $\ket{\psi_\alpha(t)}$ evolves stochastically according to either
\begin{itemize}
\item $\ket{\psi_\alpha(t+\delta t)} \propto c_\mu\ket{\psi_\alpha(t)}$ with probability $\delta p_\mu = \braket{\psi_\alpha(t)|\Gamma_\mu c_\mu^\dag c_\mu|\psi_\alpha(t)}\delta t$, or
\item $\ket{\psi_\alpha(t+\delta t)} \propto e^{-iH_\eff \delta t} \ket{\psi_\alpha(t)}$ with probability $1-\sum_\mu \delta p_\mu$.
\end{itemize}
We choose $\delta t$ so that $\delta p=\sum_\mu \delta p_\mu \ll1$.
Since this process is stochastic, we average over a sufficiently large number $N_\text{traj}$ of trajectories to estimate the values of observables:
\begin{equation}
\braket{\hat\O}=\tr[\hat\O\rho(t)] \simeq \frac{1}{N_\textrm{traj}} \sum_{\alpha=1}^{N_\textrm{traj}} \braket{\psi_\alpha(t)|\hat\O|\psi_\alpha(t)}.
\end{equation}
For the numerical data presented in the main text, we average over up to $N_\text{traj} = 1000$ trajectories, and statistical uncertainties are estimated using the bootstrapping technique~\cite{bootstrap}.
For relatively large system sizes $n>8$, numerical computations of exact many-body wavefunctions are impractical.
Instead, we store the wavefunction in an MPS representation, and simulate the evolution using the time-evolving block decimation algorithm~\cite{Vidal,*Vidal2}.
Since we are dissipatively preparing AKLT states that have bond dimension $D=2$, we find that restricting the maximum bond dimension of our MPS wavefunction to $D\le 15$ is sufficient, as truncation errors are found to be $< 5\times 10^{-5}$ in all simulations.
This algorithm allows us to simulate systems with up to $n=25$ spins.

During our simulated evolution, we monitor two observables: (1) energy density  $\braket{H_\textrm{AKLT}}/(n-1)$, and  (2) fidelity of state preparation $\mathcal{F}=\braket{P_G}$ where $P_G$ is the projector onto AKLT ground states.
Note that since our simulation in the main text is for the open boundary condition, there are four degenerate ground states that are all accepted as output; we define our fidelity $\F$ to be the sum of overlap with each accepted state.
Another widely used measure on quantum states, trace distance~\cite{TraceDistance}, cannot be applied in this context, because it measures how close a state is to another (target) state, not to a subspace of such states.
Even in a situation where the target state is a single pure state, e.g. under periodic boundary condition, the trace distance $T$ is bounded by our fidelity $\F$ through $1-\sqrt{\F}\le T \le \sqrt{1-\F}$.
In our numerics where the system size goes up to $n=25$, the computational cost of calculating the density matrix and trace distance would also be prohibitively expensive.
Finally, this fidelity $\mathcal{F}$ coincides with the success probability of state preparation, which is a physically meaningful metric.

\section{Effect of imperfect quantum jump detection\label{sec:imperfectQJdetect}}

In realistic experiments, the detection of quantum jumps often entails imperfections.
The presence of such imperfections affects our protocol by (i) not heralding the failure of connection of two chains (false-positive), and (ii) incorrectly heralding failure when the connection has been successful (false-negative).
The former may arise due to imperfect detection efficiency, and the latter due to the dark counts in the detector.
As mentioned in the main text, our parallelized protocol can still have an efficient scaling even when such imperfect quantum jump detection is accounted for.
In the false-negative scenario, we can still discard affected particle pairs and continue the procedure, but we have to adopt an ideal case success probability lower than $p_\text{max}$ (discussed below).
To minimize the occurrence of false-positives, we propose two methods that address detector inefficiency.
In the following analysis, let us denote the detector efficiency by $1-\eta$, the dark count rate by $r$, and the ideal case success probability by $p$.

\textbf{Success probability after false-negatives}---
We showed previously in Sec.~\ref{sec:connectAKLT} that the maximum success probability of connection of $p_\text{max}=1/2$ can be recovered for the subsequent attempt after a failed connection (via detection of quantum jumps), if we discard the affected particles and apply a global $\pi$-rotation $U=e^{i\pi S_y}$ to one of the remaining chains.
However, this is not the case if the quantum jump detector has only received a dark count, i.e. the failure is a false-negative, and the connection has in fact succeeded.
After a dark count is registered, the experimenter will unwittingly discard the particles at the original interface, apply $U$, and retry the connection anyways.
If we consider the density matrix of the remaining pair of chains, we can see that their edge states at the interface are essentially randomized in a maximally mixed state, which would intuitively yield a success probability $p\approx1/4$ in the attempt to connect them.
More precisely, if $k$ particles were discarded in each original chain of length $n$ starting from interface (i.e. $2k$ particles in the middle of the connected chain of length $2n$), the success probability of connecting the two chains of length $n-k$ can be computed to be
\begin{equation}
p = \frac{1}{4}(1-\epsilon^{2k})+\O(\epsilon^{n}),
\end{equation}
where $\epsilon=-1/3$. In particular, when $k=1$, $p\simeq 2/9$.

\begin{figure}[h]
\centering
\includegraphics[width=0.45\textwidth]{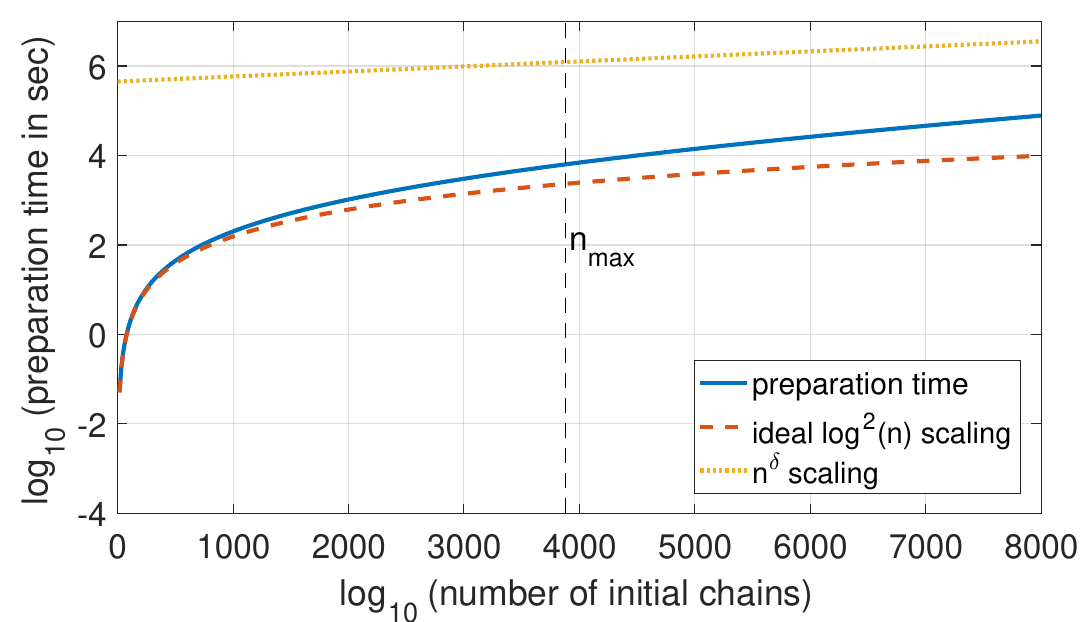}
\caption{\label{fig:ImperfectScaling}
Preparation time in the parallelized protocol with detector efficiency $1-\eta=0.2$ and dark count rate $r=25$~Hz, using jump operators of the form $c=\ketbra{++}$ (Method 1). We also assume a quantum jump rate $\tau_0^{-1}=1$~MHz, ideal case success probability $p=2/9$, time to discard atoms and reset edges $\tau_r=\tau_0$, and target final error $\E=10^{-4}$.}
\end{figure}

\textbf{Method 1 to mitigate false-positives}---
The first method to address false-positives from detector inefficiency is to only use jump operators of the form $c=\ketbra{++}$ for the connection.
Once a quantum jump occurs, the state will continue to undergo quantum jumps indefinitely, creating a much larger signal and effectively larger detection efficiency.
Consequently, the detector inefficiency can be exponentially suppressed by the time $\tau_c$ of having the jump operators turned on.
Let $\tau_0$ be the time-scale in which a single quantum jump would occur.
In this case, the probability of diagnosing a successful connection and keeping the result is given by the probability of not detecting any quantum jumps over time $\tau_c$,
\begin{align}
p_\text{succ} &= \Pr\{\text{keep}\} = p(1-r\tau_0)^{\tau_c/\tau_0} + (1-p)\eta^{\tau_c/\tau_0} \nonumber \\
&= p(1-r\tau_0)^{\tau_c/\tau_0} ( 1+ a \tilde{\eta}^{\tau_c/\tau_0}),
\end{align}
where we defined $a=(1-p)/p$ and $\tilde{\eta}=\eta/(1-r\tau_0)$.
The fidelity is the conditional probability that the diagnosed success was a truly successful connection
\begin{align}
\F &= \Pr\{\text{success}|\text{keep}\} =  \frac{p(1-r\tau_0)^{\tau_c/\tau_0}}{p(1-r\tau_0)^{\tau_c/\tau_0} + (1-p)\eta^{\tau_c/\tau_0}} \nonumber \\
&= \frac{1}{1+a\tilde{\eta}^{\tau_c/\tau_0}}
\end{align}
Note that we can only achieve a fidelity arbitrarily close to 1 if $\tilde{\eta}<1$, i.e. when the detector efficiency is larger than the dark count probability $1-\eta >r\tau_0$.
In order to achieve a final error of $\E$ for a system size of $n$ from initial chains of length $n_0$, where $n/n_0$ connections are necessary, we need $1-\F\le n_0\E/n$, and consequently $\tau_c = \O(\ln(n/n_0\E)/\ln\tilde\eta^{-1})$.
This is consistent with the $\tau_c\sim \log n$ scaling necessary in the ideal protocol.
Nonetheless, our new success probability now decreases with system size $n$ as $p_\text{succ} \sim \O(n^{-\delta})$ if dark counts are non-negligible, where $\delta=\ln(1-r\tau_0)/\ln \tilde{\eta} \approx r\tau_0/\ln\eta^{-1}$.
Consider now the average time to prepare a chain of length $n$:
\begin{equation}
T(n) = T_0 + \frac{\tau_c + \tau_r (1-p_\text{succ})} {p_\text{succ}} \log_2 \frac{n-n_c}{n_0-n_c}
\end{equation}
where $n_c=2(1-p_\text{succ})/p_\text{succ}$, $T_0$ is the time to prepare initial length-$n_0$ chains, and $\tau_r$ is some constant time necessary to reset the edge states in the event of failure.
At first sight, this indicates that our preparation time would ultimately scale polynomially instead of polylogarithmically in the infinite $n$ limit.
However, in the regime of $r\tau_0\ll 1$, this polynomial dependence has a very small power, and its effect can be neglected if $r\tau_c\ll 1$.
Hence, in practice, our protocol has an efficient, polylogarithmic scaling up to $n \ll n_\text{max} =\O((1/\eta)^{1/r\tau_0})$, beyond which it switches to a polynomial scaling.
For instance, even if the single-photon detection efficiency is $1-\eta=0.2$, then assuming a dark count rate of $r=25$~Hz~\cite{SPADspecs}  and a quantum jump scattering rate of $\tau_0^{-1}\approx 1$~MHz, it takes an astronomically long chain of $n_\text{max}\sim 10^{4000}$ to reach the polynomial scaling.
An example scaling under these conditions is shown in Fig.~\ref{fig:ImperfectScaling}.

\textbf{Method 2 to mitigate false-positives}---
The second method for addressing detector inefficiency is to slowly turn on jump operators in the vicinity of the interface.
In this way, the absence of quantum jumps further confirms that the two chains have indeed been successfully connected;
since only a successful connection does not lead to any subsequent quantum jumps, any false-positive diagnosis of successful connection can be corrected.
More concretely, consider a $k$-step scheme where we turn on jump operators to include $k$ neighbors on each side of the original interface, one pair of neighbors at a time.
At step $\ell=1,\ldots,k$, we have jump operators on for $2\ell$ particles centered at the interface, turned on for time $\tau^c_\ell$.
If a quantum jump occurred and evaded detection at any step $\ell$, we assume $\tau^c_\ell$ is long enough so that the $2\ell$ particles would have formed a connected chain of length $2\ell$.
At the subsequent step $\ell+1$, the length-$2\ell$ chain in the middle can be connected to the two length-$(n-\ell)$ chains on both sides if we succeed by having no quantum jump, producing a fully connected chain of length $2n$.
Note that the success probability for steps $\ell>1$ is roughly $p_2=1/2^4$.
From the system size scaling found in our numerical simulations presented in the main text, we expect to need $\tau^c_\ell\sim (2\ell)^{2.97}\approx (2\ell)^3$.
Additionally, we expect the number of quantum jumps during $\tau^c_\ell$ of step $\ell$ to roughly scale as $N^{\text{jump}}_\ell \approx C\tau^c_\ell$ for some constant $C$.
Observe that this scheme allows us to obtain a fully connected chain even in the event of initial failure(s), as long as we do not have any quantum jumps at the last step.
Thus, the probability of succeeding and deciding to keep the result (due to not detecting any quantum jump) is
\begin{align}
&\Pr\{\text{success and keep}\} \nonumber \\
=~& p e^{-r\sum_{\ell=1}^k \tau^c_\ell} + (1-p)\eta^{C\tau^c_1}p_2e^{-r\sum_{\ell=2}^k\tau_\ell^c} \nonumber\\
 & \quad + (1-p)\eta^{C\tau^c_1}(1-p_2)\eta^{C\tau^c_2}p_2 e^{-r\sum_{\ell=3}^k\tau_\ell^c} + \cdots \nonumber \\
=~& p e^{-rT_k}\left(1 + \frac{1-p}{p}p_2\sum_{s=1}^{k-1} (1-p_2)^{s-1} (\eta^C e^{r})^{T_s} \right),
\end{align}
where $T_s = \sum_{\ell=1}^s \tau^c_\ell\approx Bs^4$ for some constant $B$. The probability of failing at the last step but still keeping the result is
\begin{equation}
\Pr\{\text{fail and keep}\} = (1-p)(1-p_2)^{k-1}\eta^{CT_k}.
\end{equation}
The fidelity is the conditional probability of true success given that we have kept the result:
\begin{align}
\F &= \Pr\{\text{success} | \text{keep}\}  = \textstyle{\frac{ pe^{-rT_k}(1+\cdots)}{pe^{-rT_k}(1+\cdots) + (1-p)(1-p_2)^{k-1}\eta^{CT_k}}} \nonumber \\
&\ge \frac{pe^{-rT_k}}{pe^{-rT_k} + (1-p)(1-p_2)^{k-1}\eta^{CT_k}} \nonumber \\
&\approx 1- b(1-p_2)^{k-1}\xi^{k^4},
\end{align}
where we defined $b=(1-p)/p$ and $\xi=(\eta^{C}e^{r})^B$. To achieve arbitrarily good fidelity, we require $\xi<1$, i.e. the dark count rate $r < C\ln \eta^{-1}$ needs to be sufficiently small.
At the same time, the apparent ``success'' probability of keeping the result is $p_\text{succ}=\Pr\{\text{keep}\}\approx pe^{-rBk^4}$.
We can carry out the same analysis as in the previous method, and a similar behavior emerges: when the dark count rate is nonzero, the efficient polylogarithmic scaling applies until a maximum chain length of $n\ll n_\text{max} = \O(\eta^{-C/r})$, beyond which a polynomial scaling of $\O(n^{\delta'})$ with $\delta'\approx r/\ln\eta^{-C}$ applies.

\section{Analysis of Rydberg-EIT implementation proposal\label{sec:detail_REIT}}
In this appendix, we derive the effective dissipative interaction between two nearby particles for our Rydberg-EIT implementation scheme introduced in the main text.
Consider two particles interacting via the Rydberg shift $H_\text{int} = U\ketbra{rr}$. Their effective (non-Hermitian) Hamiltonian under the Rydberg-EIT scheme proposed in the main text is
\begin{eqnarray}
H_\eff &=& \sum_{j=1}^2\Big[(g\ketbrat{+}{e} + \Omega\ketbrat{r}{e} + h.c.) - i\frac{\gamma}{2}\ketbra{e}\Big]_j  + H_\text{int} \nonumber \\
&=& \sum_{j=1}^2\Big[\Delta (\ketbrat{B}{e}+ h.c.) - i\frac{\gamma}{2}\ketbra{e}\Big]_j + H_\text{int},
\end{eqnarray}
where $\Delta=\sqrt{\Omega^2+g^2}$, $\ket{D} = (\Omega\ket{+}-g\ket{r})/\Delta$ is the EIT-dark state, and $\ket{B} = (g\ket{+} + \Omega\ket{r})/\Delta$ is a state orthogonal to $\ket{D}$ that we call EIT-bright state. Now consider a general two-particle wavefunction $\ket{\psi} = \sum_a c_{aa}\ket{aa} + \sum_{a<b} c_{ab} (\ket{ab}+\ket{ba})/\sqrt{2}$, where we have restricted ourselves to working in the symmetric subspace.
Then the equations of motion for the coefficients are
\begin{align*}
i\dc_{DD} &= \frac{g^4U}{\Delta^4} c_{DD} - \frac{\sqrt{2}g^3\Omega U}{\Delta^4} c_{DB} + \frac{g^2 \Omega^2 U}{\Delta^4} c_{BB}, \\
i\dc_{De} &= -i\frac{\gamma}{2}c_{De} + \Delta c_{DB}, \\
i\dc_{DB} &= {\textstyle \frac{2g^2\Omega^2 U}{\Delta^4} c_{DB} + \Delta c_{De} - \frac{\sqrt{2}g^3\Omega U}{\Delta^4} c_{DD} - \frac{\sqrt{2}g\Omega^3 U}{\Delta^4} c_{BB} },\\
i\dc_{ee} &= -i\gamma c_{ee} + \sqrt{2}\Delta c_{eB}, \\
i\dc_{eB} &= -i\frac{\gamma}{2} c_{eB} + \sqrt{2}\Delta (c_{ee} + c_{BB}), \\
i\dc_{BB} &= {\textstyle \frac{\Omega^4 U}{\Delta^4} c_{BB} + \sqrt{2}\Delta c_{eB}  - \frac{\sqrt{2}g\Omega^3 U}{\Delta^4} c_{DB} + \frac{g^2\Omega^2 U}{\Delta^4} c_{DD} }.\\
\end{align*}
In the limit of $U\ll g,\Omega,\gamma$, or $g\ll \Omega,\gamma,U$, and assuming we start initially with $\ket{\psi}=\ket{DD}$, we can adiabatically eliminate the fast dynamics involving coefficients $\{c_{ab}\}$ other than $c_{DD}$. This procedure can be effectively achieved by setting $\dot{c}_{ab}=0$ for ${ab}\neq DD$, allowing us to obtain
\begin{gather}
i\dc_{DD} = U_{DD} c_{DD} \quad \text{where} \quad
U_{DD}
= \frac{g^4}{\Delta^4} \frac{U}{1+i\chi U}
\nonumber \\
\text{and}\quad \chi = \frac{\Omega^2[\Omega^2+(1+3g^2/\Delta^2)\gamma^2/4]}{\Delta^4\gamma}.
\end{gather}
Here, $\Re[U_{DD}]$ is the interaction-induced energy shift, and $\Gamma_{DD}=-2\Im[U_{DD}]$ is the two-body effective decay rate.

A more general version of adiabatic elimination for open system can be found in Ref.~\cite{EffOpOpenSys}, which allows us to obtain effective jump operators.
Consider original jump operators of the form $L_{s,j} = \ketbrat{s}{e}_j$, corresponding to the spontaneous decay from excited state $\ket{e}$ to one of the three hyperfine ground state $\ket{s}$ for $s\in\{+,0,-\}$ in atom $j$.
Note that in practice, the excited state can also decay into other hyperfine ground states, which can then be repumped to the excited state using additional lasers.
We denote the decay rate corresponding to $L_{s,j}$ by $\gamma_s$, where $\gamma_+ + \gamma_0 + \gamma_-=\gamma$.
Then we can compute the effective jump operators:
\begin{equation}
L^\eff_{s,1} = \ketbrat{s\tilde{e}}{DD}, \quad L^\eff_{s,2} = \ketbrat{\tilde{e}s}{DD},
\end{equation}
with rate $\Gamma_\eff = \frac{\gamma_s}{2\gamma}\Gamma_{DD}$, and $\ket{\tilde{e}}\propto \frac{\Omega \gamma}{2\Delta}\ket{B} - i\Omega\ket{e} - \frac{g\gamma}{\Delta}\ket{D}$.
Note that $\ket{\tilde{e}}$ will further decay through the original jump operator $L_{s,j}$. Assuming we are in the regime $\gamma \gg \gamma_s\Gamma_{DD}/2\gamma$ so that $\ket{\tilde{e}}$ is a short-lived intermediate state, we can approximate the overall effective dynamics with jump operators of the form $L^\eff_{ss'}=\ketbrat{ss'}{DD}$ for $s,s'\in\{+,0,-\}$.
While we only need a jump operator such as $L^\eff_{00}=\ketbrat{00}{DD}$ to ensure AKLT states are unique steady states of the engineered dissipative dynamics, the additional effective jump operators do not affect the steady states and can in fact help to more quickly depopulate the undesired states.

\section{Scaling of imperfection in Rydberg-EIT implementation\label{sec:imperfection-scaling}}
Our protocol prepares AKLT states with finite fidelity when experimental imperfections are taken into account.
Here, we analyze how the fidelity scales as multiple chains are connected.
In particular, the long-range nature of interaction in the proposed Rydberg-EIT implementation limits fidelity even in the absence of dephasing.
Nevertheless, we show that by adopting the parallelized protocol, the long-range interactions only affect the initial preparation of length-$n_0$ chains, and such imperfection does not substantially grow in the later connection procedures.
This is because the connections involve turning on the dissipative interaction on particle pairs that are spatially separated by at least $n_0$ particles.
Since the effective decay rate scales as $\Gamma_{DD}\sim1/R^{12}$ for the proposed implementation, the perturbative effect of long-range interaction is characterized by the very small number of $1/(n_0-1)^{12}$, which becomes even smaller in later rounds of connections.
Hence, we neglect the effect of long-range interaction on the connections, and only consider how the induced errors on the states of initial chains propagate through the protocol.
Let us assume that we initially start with individual chains of length $n_0$, each with bounded error $\epsilon_0$.
At the $\ell$-th level of connections, we on average double the length $n_\ell\simeq 2n_{\ell-1}-n_c$, where $n_c$ is the expected number of particles discarded in each connection.
The number of initial chains necessary to reach a final chain length of $n$ is $L=(n-n_c)/(n_0-n_c)$, and $L-1$ connection procedures need to be performed.
Hence, the final error is bounded by
\begin{equation}
1-\F \le L\epsilon_0 + (L-1)\times \O(e^{-\gamma_1\tau_c}) \approx n\epsilon_0/n_0,
\end{equation}
where we neglect the second term which can be made small compared to the first if we choose $\tau_c=\O(\ln n)$.
As we can see, the predominant source of error is due to the imperfect initial chains, whose errors add linearly.

This linear scaling of error is indeed very favorable for a many-body state preparation protocol.
To put this in perspective, let us estimate how the effective temperature $T_\eff$ scales in our connection procedure. We define the effective temperature through the relation
\begin{align}
\F &= \tr\left[P_G\frac{e^{-H/T_\eff}}{\tr[e^{-H/T_\eff}]}\right] = \frac{1}{1+\int_{\Delta_\textrm{gap}}^\infty \rho(E) e^{-E/T_\eff} dE } \nonumber \\
&\approx 1-\int_{\Delta_\textrm{gap}}^\infty \rho(E) e^{-E/T_\eff} dE ,
\end{align}
where $\rho(E)$ is the density of states at energy $E$, and we assume that $T_\eff$ is sufficiently small.
Now we consider connecting two length-$n_1$ chains at effective temperature $T_1$ (with errors $1-\F_1$).
After connection, we have a chain of length $n_2\approx 2n_1$, with bounded error $1-\F_2\lesssim 2(1-\F_1)$.
The corresponding effective temperature $T_2$ of the connected chain can be estimated from
\begin{align}
&\int_{\Delta_\textrm{gap}}^\infty \rho_2(E) e^{-E/T_2} dE \approx 1-\F_2 \nonumber \\
& \qquad \lesssim 2(1-\F_1) \approx \int_{\Delta_\textrm{gap}}^\infty 2\rho_1(E) e^{-E/T_1} dE 
\end{align}
which implies
\begin{align}
 \int_{\Delta_\textrm{gap}}^\infty 2\rho_1(E) e^{-E/T_1} dE  -
 \int_{\Delta_\textrm{gap}}^\infty \rho_2(E) e^{-E/T_2} dE
 \gtrsim 0,
\end{align}
where $\rho_1(E)$ and $\rho_2(E)$ denote the density of states for chains of length $n_1$ and $n_2 \approx 2 n_1$, respectively.
In a generic many-body interacting system, the density of states grows exponentially in system sizes.
Here, we are most interested in the density of states of low-lying excitations, e.g. the first excited band, where the scaling of $\rho(E)$ can be much weaker.
Ref.~\cite{ArovasMagnon1988} used a Bijl-Feynman single-mode approximation to deduce that there is a band of low-lying excited states with dispersion relation $E_1(k)=\frac{5}{27}(5+3\cos k)$, corresponding to magnon excitations.
Therefore, we expect the number of states in the low-lying excited bands to scale at least linearly with system size, and thus $\rho_2(E)\ge 2\rho_1(E)$.
Applying this to the earlier inequality, we have
\begin{gather}
\int_{\Delta_\textrm{gap}}^\infty \rho_2(E) (e^{-E/T_1}-e^{-E/T_2}) dE \gtrsim 0 \nonumber \\
~ \Longrightarrow ~
T_2 \lesssim T_1.
\end{gather}
Hence, the effective temperature should not increase (and can potentially decrease) after each connection procedure.

\section{Generalization to symmetric MPSs\label{appx:generalization}}

In this appendix, we provide additional details on how to generalize our protocol for a broader class of translation-invariant MPSs with internal symmetry.
We first introduce notation and elaborate on a few useful properties (including injectivity) of translation-invariant MPSs in Sec.~\ref{sec:notations}.
More detailed descriptions and proofs of these properties can be found in Ref.~\cite{MPSreview}.
We then discuss the meaning of internal symmetry of MPSs in Sec.~\ref{sec:InternalSymmetry}.
Finally, we describe and analyze the generalization of our protocol in Sec.~\ref{sec:general}, and prove the lower bound for the minimum complexity of decay channels.

\subsection{Notations and relevant properties of matrix product states\label{sec:notations}}
Any (unnormalized) translation-invariant MPS with physical dimension $d$ (i.e. spin-$\frac{d-1}{2}$ particles) and bond dimension $D$ can be written as:
\begin{equation}
\ket{A_{ab}^n} = \sum_{\{s_i\}} \braket{a|A^{(s_1)}A^{(s_2)}\cdots A^{(s_n)}|b}\ket{s_1s_2\cdots s_n},
\end{equation}
where $s_i\in\{1,2,\ldots,d\}$ runs over the physical spin basis for the $i$-th particle, and $\ket{a},\ket{b}\in\C^D$ indicate the ``boundary conditions''.
We denote by $\ket{A^n_{ab}}$ the translation-invariant MPS of $n$ particles with open boundary condition specified by $a$ and $b$, and by $\ket{A_\circ^n}$ the MPS of an $n$-particle system with periodic boundary condition, i.e. $\ket{A_\circ^n} = \sum_a \ket{A_{aa}^n}$.
Under open boundary condition, there could be at most $D^2$ distinct states with different possible boundary conditions, e.g. four-fold degeneracy of AKLT states.
In general, however, these $D^2$ states may not be linearly independent unless the MPS is \emph{injective} (defined below).

\textbf{Canonical Form}---
An MPS is in a \emph{canonical form} if the matrices have a common block diagonal structure: $A^{(s)}=\textrm{diag}(\lambda_1A_1^{(s)}, \ldots, \lambda_B A_B^{(s)}) = \bigoplus_{\beta=1}^B \lambda_\beta A_\beta^{(s)}$, where $0<\lambda_\beta\le 1$ for each block $\beta\in \{1,\ldots,B\}$.
The matrices in each block must satisfy the conditions that
(i) $\sum_s A^{(s)}_\beta A^{(s)\dag}_\beta =\I$,
(ii) a map defined as $\E_\beta(X)=\sum_s A_\beta^{(s)}X A_\beta^{(s)\dag}$ has $\I$ as its only fixed point (unique eigenvector with unity eigenvalue),
and finally (iii) $\sum_s A_\beta^{(s)\dag}\Lambda_\beta A_\beta^{(s)} = \Lambda_\beta$ for some diagonal positive and full-rank matrices $\Lambda_\beta$.
From now on, we assume any MPS under consideration is written in a canonical form.

\textbf{Transfer Matrix}---
Consider the completely positive map $\E(X)=\sum_s A^{(s)}X A^{(s)\dag}$, or equivalently the \emph{transfer matrix} $\T=\sum_s A^{(s)*}\otimes A^{(s)}$.
Understanding the spectrum of this transfer matrix is useful for computing the expectation value of an observable or the overlap between two quantum states, e.g. $\braket{A_{ab}^n|A_{a'b'}^n} = \braket{aa'|\T^n|bb'}$~\cite{MPSreview}.
Some known eigenvectors of $\T$ are $\frac{1}{\sqrt{D_\beta}}\sum_{i\in\beta}\ket{ii}$, where $D_\beta$ is the dimension of the $\beta$-th block, and correspond to eigenvalues $|\lambda_\beta|^2$.
Denoting the other eigenvectors of $\T$ with eigenvalues $\epsilon_\nu$ by $\ket{\nu}$, we have
\begin{widetext}
\begin{equation} \label{eq:genMPSoverlap}
\braket{A_{ab}^n|A_{a'b'}^n} = \braket{aa'|
\left[\sum_\beta \frac{|\lambda_\beta|^2}{D_\beta}\sum_{i,j \in\beta}\ketbrat{ii}{jj}
+ \sum_\nu \epsilon_\nu \ketbra{\nu}\right]^n
\hspace{-5pt} |bb'}
 = \sum_\beta
\frac{|\lambda_\beta|^{2n}}{D_\beta}\delta_{a,b\in\beta}\delta_{aa'}\delta_{bb'} + \sum_\nu\epsilon_\nu^{n}\braket{aa'|\nu}\braket{\nu|bb'}.
\end{equation}
\end{widetext}
Since $|\lambda_\beta|^2$ is the largest eigenvalue of each block $\beta$, typically only the first term is relevant in the limit of large $n$.

\textbf{Parent Hamiltonian}---
For a sufficiently large $L$,  the set of matrix products $\{A^{(s_1)}\cdots A^{(s_L)}: 1\le s_i\le d\}$ spans the vector space of all matrices with the same block diagonal structure as the canonical form~\cite{MPSreview}.
We call $L$ the \emph{interaction length} of the MPS.
Without loss of generality, we can assume that $L=2$.
This is because otherwise we can group $L$ sites together to get an equivalent MPS with larger physical dimension $d'\le d^L$, and a new interaction length $L'=2$.
The \emph{parent Hamiltonian} of an MPS is then defined to be $H_p=\sum_i h^{(i)}$, where $h$ is any positive semi-definite operator acting on nearest neighboring sites, whose kernel is
\begin{equation}
\ker(h) = \text{span}\{\ket{A_{ab}^2}: \forall a,b\}.
\end{equation}
In other words, $H_p$ imposes a condition $h^{(i)}$ for every pair of neighboring sites $(i,i+1)$, which our MPS trivially satisfies (i.e. $h^{(i)}\ket{A_{ab}^n}=0$ for all $1\le i\le n-1$).
Hence, $H_p$ is a frustration-free Hamiltonian of which the MPS is a zero-energy ground state.
The ground-state degeneracy depends on both the boundary condition and the number of blocks in the MPS canonical form.
Ref.~\cite{MPSreview} has shown that under periodic boundary condition, this degeneracy equals the number of independent blocks in the canonical form, since the ground space consists of MPSs constructed from the sub-matrices from every block.

\textbf{Injectivity}---
Often it is useful to assume a condition that the MPS is \emph{injective}, which is satisfied in the generic case except for specific, fine-tuned MPSs~\cite{MPSreview}.
This injectivity condition is that the transfer matrix $\T$ has only one eigenvector corresponding to its largest eigenvalue (which we normalize to 1 in the canonical form).
This also implies that there is just one block in the canonical form of the MPS.
In this case, Eq.~\eqref{eq:genMPSoverlap} simplifies to
\begin{equation} \label{eq:Overlap}
\braket{A_{ab}^n|A_{a'b'}^n} = \braket{aa'|\T^n|bb'} = \frac{1}{D}\delta_{aa'}\delta_{bb'} + \O(\epsilon_2^n),
\end{equation}
where $\epsilon_2$ is the second largest eigenvalue of $\T$.
Additionally, this implies that the parent Hamiltonian under periodic boundary condition has the MPS as its unique ground state, and that the ground state energy is gapped in the thermodynamic limit.
Under open boundary condition, $D^2$ distinct boundary conditions give rise to $D^2$ linearly independent and degenerate ground states $\ket{A^n_{ab}}$.
By appropriately modifying the parent Hamiltonian terms at the boundaries, we can break the degeneracy and make one of the $D^2$ states the unique ground state.

\subsection{Internal Symmetries of MPSs\label{sec:InternalSymmetry}}
We say a translation-invariant MPS defined on $d$-dimensional physical spins respects an internal symmetry $\cG$, if for some unitary representation $U:\cG\to U(d)$, we have:
\begin{align}
&U_g^{\otimes n} \ket{A_{ab}^n} = \sum_{a'b'}[\chi_g]_{ab}^{a'b'} \ket{A_{a'b'}^n}
\nonumber \\
\text{and} \quad 
&U_g^{\otimes n} \ket{A_\circ^n} = e^{i\theta_g}\ket{A_\circ^n}.
\end{align}
That is, a global action of the symmetry operation keeps a ground state of the MPS parent Hamiltonian in the ground space under open boundary condition, or only imprints a complex phase factor under periodic boundary condition.

Assuming the symmetry group is reasonable (either a discrete or a compact connected Lie group), but without assuming injectivity, Ref.~\cite{MPSsymmetry} showed that we can replace the action of the symmetry in the physical basis with a unitary in the virtual bond basis. More explicitly, we have
\begin{equation}
\sum_{s'} [U_g]_{ss'} A^{(s')} = w_g u_g A^{(s)} u_g^\dag.
\end{equation}
Here, $u_g = P_g v_g$, with $v_g=\bigoplus_{\beta=1}^B v_g^\beta$ taking on the same block diagonal structure as $A^{(s)}$, and each $v_g^\beta$ a unitary in block $\beta$. $P_g$ is a permutation amongst the $B$ blocks. Lastly, $w_g = \bigoplus_\beta e^{i\varphi_g^\beta}\I_\beta$ is a phase factor for each block. If $\cG$ is a compact connected Lie group, Ref.~\cite{MPSsymmetry} showed that $P_g=\I$, while $g\mapsto e^{i\varphi_g^\beta}$ and $g\mapsto v_g^\beta$ are representations of $\cG$.

For AKLT states, where $\cG= \SO(3)$, the relevant representation is given by the rotations $U_g = \exp(i\vec\alpha_g\cdot\vec{S})$ on the spin-1 vector $\vec{S}$ for some real parameters $\vec\alpha_g=(\alpha_g^x,\alpha_g^y,\alpha_g^z)$.
In particular, we have
\begin{align}
&\sum_{s'} [U_g]_{ss'} A^{(s')} = u_g A^{(s)} u_g^\dag \nonumber \\
 \Longrightarrow \quad 
& U_g^{\otimes n} \ket{A_{ab}^n} = \ket{A_{u_g^\dag a,u_g^\dag b}^n},
\end{align}
where $u_g = \exp(i\vec{\alpha}'_g\cdot\vec\sigma/2)$, with $\vec{\alpha}'_g=(\alpha_g^x,-\alpha_g^y,\alpha_g^z)$.

\subsection{Finding a minimal set of decay channels using symmetry\label{sec:general}}
Our goal is to find a minimal set $k_\textrm{min}$ of decay channels $\{ c_1, c_2, \dots, c_{k_\textrm{min}} \}$ acting on neighboring pairs of particles that deterministically prepare the desired MPS, assuming global symmetry operations are available.
For concreteness we will first focus on decay channels of the form $c_\mu=\ketbrat{\phi_\mu}{\psi_\mu}$.
We show there is a lower bound on $k_{\min}$ from the structure of representation of $\cG$ on the physical particles, and provide a construction of the jump operators saturating the bound.
The uniqueness of steady states under the constructed jump operators can be analytically confirmed using the same inductive proof strategy demonstrated for the case of AKLT states.

Without loss of generality, we assume that the desired states are ground states of a gapped, frustration-free parent Hamiltonian $H_p =\sum_i h^{(i)}$, where $h^{(i)}$ is a translation-invariant, nearest-neighbor projector that respects the internal symmetry $\cG$~\cite{MPSreview,MPSsymmetry}.
We note the projector $h$ has a block diagonal form, corresponding to different irreducible representations of $\cG$.
We call the two-particle subspace that $h$ projects onto a ``bright manifold'' $\B \equiv \range(h) \subset \mathds{C}^{d^2}$.
The ground states are uniquely characterized by vanishing populations in $\B$ for every neighboring pair of particles.

The foremost necessary condition for the jump operators $\{c_\mu\}$ is
\begin{gather} 
\B = \range\left(\sum_{\mu=1}^{k_\text{min}} Q_\mu\right)
\quad \text{where} \nonumber \\
Q_\mu  = \frac{1}{|\cG|}\sum_{g\in\cG} V_g^\dag c_\mu^\dag c_\mu V_g = \frac{1}{|\cG|}\sum_{g\in\cG} V_g^\dag \ketbra{\psi_\mu} V_g.
\label{eq:jumpcondition}
\end{gather}
In other words, the jump operators must be capable of depopulating the entire bright manifold after averaging over all symmetry operations. While here we have assumed that the symmetry group $\cG$ is finite for simplicity, the following results apply to any compact group by replacing the sum over $g\in \cG$ by an integral over the Haar measure of $\cG$.

To find the minimum number $k_\text{min}$ of $\ket{\psi_\mu}$ (and consequently $c_\mu$) required, it is useful to decompose $V_g$ into direct sums of irreducible representations (irreps) $V_g = \bigoplus_r V_g^r$, where $r$ enumerates the irreps, each with dimension $d_r$.
This decomposition is possible because finite-dimensional unitary representations of any group are completely reducible~\cite{ArtinAlgebra}.
Let $\ket{\psi_\mu}=\bigoplus_r\ket{\psi_\mu^r}$, where each $\ket{\psi_\mu^r}$ is a $d_r$-dimensional vector.
Observe that for any $d_r\times d_{r'}$ matrix $X$, we can derive the following identity using Schur's lemma~\cite{ArtinAlgebra}:
\begin{equation}
\frac{1}{|\cG|}\sum_{g\in\cG} V_g^r X V_g^{r'\dag} =
\begin{cases}
0 & \textnormal{ if } r\not\simeq r' \\
\frac{\tr(U_{rr'}^\dag X)}{d_r} U_{rr'} & \textnormal{ if } r\cong r'
\end{cases},
\end{equation}
where $r\cong r'$ means $V_g^{r} = U_{rr'} V_g^{r'} U_{rr'}^\dag$, or $r$ is equivalent (isomorphic) to $r'$ up to a unitary basis change.
Note we can always choose a basis for the representation of $V_g$ that absorbs $U_{rr'}$, so we assume $U_{rr'}=\I$ without loss of generality.
Using the notation $\bigoplus_{r,r'} M_{r,r'}$ to denote the matrix whose $r$-th row, $r'$-th column block is $M_{r,r'}$, we can write $Q_\mu$ through the above identity as
\begin{align}
Q_\mu &= \bigoplus_{r,r'} \frac{1}{|\cG|}\sum_{g\in \cG} V_g^r\ketbrat{\psi_\mu^r}{\psi_\mu^{r'}}^\dag V_g^{r'\dag} \nonumber\\
&= \bigoplus_{r} \frac{\I_r}{d_r}\braket{\psi_\mu^r|\psi_\mu^r} +  \bigoplus_{r\neq r',r\cong r'} \frac{\I_r}{d_r}\braket{\psi_\mu^{r'}|\psi_\mu^{r}},
\end{align}
where the second term characterizes the possible nonzero off-diagonal blocks, which can only exist between pairs of equivalent irreps.

Since inequivalent irreps are decoupled, we for now only consider the subspace $\B_r\subseteq \B$ corresponding to $K_r$ copies of irreps equivalent to irrep $r$ ($\dim\B_r = K_r d_r$).
Let $Q_\mu^r=Q_\mu|_{\B_r}$ as the operator $Q_\mu$ restricted to the subspace $\B_r$.
Observe that $\ket{\psi_\mu}$ restricted to this subspace is specified by the set of $K_r$ vectors $\{\ket{\psi_\mu^s} \in \C^{d_r}\}_{s=1}^{K_r}$.
When $K_r > d_r$, regardless of the choice of $\ket{\psi_\mu}$, there are $K_r-d_r$ linearly independent vectors $\vec{\beta}^j\in\C^{K_r}$, $j=1,\ldots,K_r-d_r$, such that $\sum_{s=1}^{K_r} \beta^j_s\bra{\psi_\mu^s}=0$.
Then any vectors of the form $\ket{\chi}=\bigoplus_{s=1}^{K_r}\beta_s^j\ket{v}$ are in the kernel of $Q_\mu^r$ for any $\ket{v}\in\C^{d_r}$, since one can verify $Q_\mu^r\ket{\chi} = 0$.
Since there are $(K_r-d_r)d_r$ linearly independent such vectors $\ket{\chi}$, we have $\rank(Q_\mu^r)\le d_r^2$.
Hence, in order to fully depopulate $\B_r$, we need $\text{dim}(\B_r) = \rank(\sum_\mu Q_\mu^r) \le \sum_\mu \rank(Q_\mu^r) \le k_\text{min} d_r^2$.
Because $k_\text{min}$ must be an integer, we must have $k_\text{min} \ge \lceil K_r/d_r\rceil$, for every irrep $r$.

Note that this lower bound for $k_\text{min}$ can be saturated by construction as follows.
First, we partition the $K_r$ equivalent irreps into $\lceil K_r/d_r\rceil$ groups of no more than $d_r$ irreps.
For each group, we can assign a $\ket{\psi_\mu}$ that is nonzero only in the subspace corresponding to the irreps in the group.
Lastly, we make all off-diagonal blocks vanish for each group $\mu\in\{1,\ldots,\lceil K_r/d_r\rceil\}$, by finding $d_r$ or fewer mutually orthogonal vectors $\ket{\psi_\mu^r}\in \C^{d_r}$ such that $\braket{\psi_\mu^{r'}|\psi_\mu^r}=0$ for $r\neq r'$.
For a single jump operator of the form $c_\mu = \ketbrat{\phi_\mu}{\psi_\mu}$,  the state $\ket{\psi_\mu}$ may have supports on more than one of subspaces $\B_r$.
Therefore, the construction of a set of jump operators $\{c_\mu\}$ to satisfy Eq.~\eqref{eq:jumpcondition} can be done in parallel for all the different $\B_r$ corresponding to the inequivalent set of irreps, leading to the minimum number
\begin{equation} \label{eq:k_lower_bound}
k_\text{min} = \max_r \lceil K_r/d_r \rceil.
\end{equation}
Here, $r$ enumerates inequivalent irreps of $\cG$ in $\B$, $K_r$ is the number of copies of $r$, and $d_r$ is the dimension of $r$.

We can also consider an arbitrary jump operator $c_\mu$ beyond the rank-1 form of $\ketbrat{\phi_\mu}{\psi_\mu}$.
For any operator $c_\mu$, we can perform singular value decomposition to write $c_\mu=\sum_{i_\mu} \sqrt{\gamma_{i_\mu}}\ketbrat{\phi_{i_\mu}}{\psi_{i_\mu}}$, where $\braket{\phi_{i_\mu}|\phi_{j_\mu}}=\braket{\psi_{i_\mu}|\psi_{j_\mu}}=\delta_{i_\mu j_\mu}$.
Then $c_\mu^\dag c_\mu = \sum_{i_\mu} \gamma_{i_\mu}\ketbra{\psi_{i_\mu}}$ with $\gamma_{i_\mu}>0$.
Hence, the condition of Eq.~\eqref{eq:jumpcondition} becomes a condition imposed on the set of right-singular vectors $\{\ket{\psi_{i_\mu}}: \forall \mu,i_\mu\}$, where we must have $\B = \range(\frac{1}{|\cG|} \sum_{g,\mu, i_\mu}\gamma_{i_\mu} V_g^\dag \ketbra{\psi_{i_\mu}} V_g)$.
We can thus interpret the $k_\textrm{min}$ found for rank-1 jump operators as the minimum number of independent $\ket{\psi_{i_\mu}}$'s.

While we can easily construct a minimal set of $\{c_\mu\}$ to satisfy the necessary condition of Eq.~\eqref{eq:jumpcondition}, we still need to prove the uniqueness of steady states.
This can be done using our inductive proof strategy, where one simply needs to confirm that there are only trivial solutions to Eq.~\eqref{eq:open-uniqueness} under open boundary conditions.
As discussed in Sec.~\ref{sec:general-main}, this simply involves showing that a certain matrix $\vect{M}$ has full rank.
Nevertheless, for non-injective MPSs, this scheme cannot break the ground-state degeneracy intrinsic to the MPS parent Hamiltonian, but it can guarantee that the ground states are the only steady states.

\bibliography{refs}

\end{document}